\documentclass[journal]{IEEEtran}
\usepackage{hyperref}
\usepackage{amsmath,amsfonts}
\usepackage{algorithm}
\usepackage{array}
\usepackage[caption=false,font=normalsize,labelfont=sf,textfont=sf]{subfig}
\usepackage{textcomp}
\usepackage{stfloats}
\usepackage{url}
\usepackage{verbatim}
\usepackage{graphicx}
\usepackage{cite}
\usepackage{cleveref}
\usepackage{tabu}
\usepackage{booktabs}
\usepackage{bm}
\usepackage{makecell}
\usepackage{mathtools}
\usepackage{algpseudocode} 
\usepackage{xspace}
\usepackage{xcolor}
\usepackage{multirow}
\DeclarePairedDelimiterX{\infdivx}[2]{(}{)}{%
  #1\;\delimsize\|\;#2%
}
\newcommand{\infdiv}{D_\text{KL}\infdivx}
\newcommand*\dif{\mathop{}\!\mathrm{d}}

\DeclareMathOperator*{\argmin}{arg\,min}
\DeclareMathOperator\erf{erf}

\newcommand{\clrb}{\textcolor{black}}

\algnewcommand{\And}{\textbf{and}\xspace}
\algnewcommand{\Or}{\textbf{or}\xspace}

\graphicspath{{figs/}{figures/}{pictures/}{images/}{./}} 

\begin{document}

\title{Improving Efficiency of Iso-Surface Extraction on Implicit Neural Representations Using Uncertainty Propagation}

\author{Haoyu Li and Han-Wei Shen
\thanks{Haoyu Li and Han-Wei Shen are with the Department of Computer Science and Engineering, The Ohio State University, Columbus, US}
}

\markboth{Journal of \LaTeX\ Class Files,~Vol.~*, No.~*, *~20**}%
{Haoyu Li, Han-Wei Shen \MakeLowercase{\textit{et al.}}: Improving Efficiency of Iso-Surface Extraction on Implicit Neural Representations Using Uncertainty Propagation}


\maketitle

\begin{abstract}
Implicit Neural representations (INRs) are widely used for scientific data reduction and visualization by modeling the function that maps a spatial location to a data value. Without any prior knowledge about the spatial distribution of values, we are forced to sample densely from INRs to perform visualization tasks like iso-surface extraction which can be very computationally expensive. Recently, range analysis has shown promising results in improving the efficiency of geometric queries, such as ray casting and hierarchical mesh extraction, on INRs for 3D geometries by using arithmetic rules to bound the output range of the network within a spatial region. However, the analysis bounds are often too conservative for complex scientific data. In this paper, we present an improved technique for range analysis by revisiting the arithmetic rules and analyzing the probability distribution of the network output within a spatial region. We model this distribution efficiently as a Gaussian distribution by applying the central limit theorem. Excluding low probability values, we are able to tighten the output bounds, resulting in a more accurate estimation of the value range, and hence more accurate identification of iso-surface cells and more efficient iso-surface extraction on INRs. Our approach demonstrates superior performance in terms of the iso-surface extraction time on four datasets compared to the original range analysis method and can also be generalized to other geometric query tasks.
\end{abstract}

\begin{IEEEkeywords}
Iso-surface extraction, implicit neural representation, uncertainty propagation, affine arithmetic.
\end{IEEEkeywords}

\section{Introduction}

\IEEEPARstart{I}{mplicit} neural representations (INRs), as a function mapping from spatial locations to scalar values, can represent both structured and unstructured data in a continuous and succinct way. These characteristics make them suitable for scientific data reduction \cite{lu2021compressive, weiss2022fast}.
Moreover, combined with the temporal information or the priors about the parameters of a simulation, INRs can be easily extended to represent spatio-temporal data or work as a surrogate model to provide efficient preview visualization for complex simulations \cite{han2022coordnet}. These advantages make INR a promising representation of scientific data. However, visualization and analysis techniques on implicit neural representations are not fully explored.

Among many visualization techniques, iso-surface extraction is a widely used technique for visualizing scalar fields. The most popular approach for iso-surface extraction, Marching Cubes\cite{lorensen1987marching}, requires that the scalar field be discretized into a grid, and triangle patches are extracted from the grid cells using linear interpolation. To ensure that the linear assumption within each grid cell holds, the reconstruction grid should be dense enough. However, when using INRs, a dense reconstruction is computationally expensive.  Although approaches to increase the INR inference efficiency by utilizing the tensor cores' on-chip memory in modern GPUs were proposed\cite{weiss2022fast, wu2022instant}, these approaches depend on specific architecture and hardware.
\clrb{One appealing method to reduce the computation is through subdividing the field hierarchically and identifying the active cells that contain a component of the iso-surface \cite{wilhelms1992octrees,sutton1999isosurface}.
However, to build the hierarchical data structure for iso-surface extraction, the minimum and maximum values in the hierarchical nodes are needed.} For INRs, there is no easy way to get the exact minimum and maximum values without dense reconstruction of the scalar values inside the regions. Preprocessing the INRs to get values is, again, computationally very expensive and will introduce non-negligible storage overhead, which defeats the purpose of using INR as an implicit, succinct, and continuous data representation.

\begin{figure}[tbp]
  \centering 
  \includegraphics[width=\columnwidth]{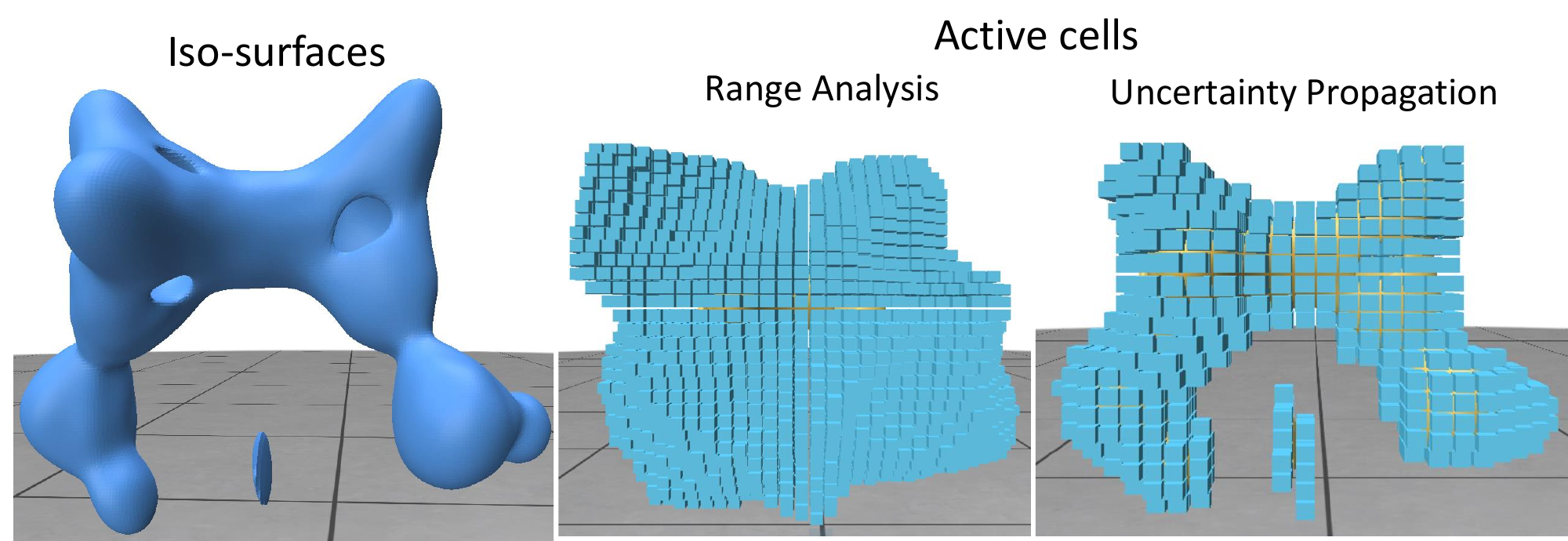}
  \caption{%
  	Experiments of the iso-surface extraction on the Ethanediol dataset. The left image shows the example iso-surface extracted using dense reconstruction and marching cubes. The middle and the right images show the active cells found by the range analysis and our approach. Our approach finds more accurate active cells with less amount of INR reconstruction.
  }
  \label{fig:preliminary}
\end{figure}

One recently proposed approach \cite{sharp2022spelunking} innovatively applies range analysis to bound the output of INR given a region of input. The output bounds are useful in many general geometric query tasks on INRs. 
Using range analysis on INR of scalar fields, we can easily build spatial hierarchies and skip regions that do not contain the chosen iso-surface to increase extraction efficiency without preprocessing or storing metadata. 
However, the coarse approximations made by range analysis do not produce exactly the same bound as the true output range of the INR. The range analysis applies affine approximations to INR and uses the arithmetic rules to guarantee that the true output range is inside the approximated bounds by the analysis. However, the estimated bound is often too large, and hence too conservative to allow effective skipping of uninterested regions. Therefore the extraction efficiency is not improved. 
Our preliminary experiments on INRs of scalar fields shown in the middle of \cref{fig:preliminary} indicate that the bounds from the range analysis are overly conservative. \clrb{A possible reason is that scalar fields from scientific simulations have more value oscillations and high-frequency features compared to signed distance functions or occupancy functions of 3D geometric models. INR learns more complex nonlinear functions leading to large errors between the affine approximation used by the range analysis and the INR}

To overcome the computation challenges of iso-surface extraction on scalar fields represented by INRs, in this paper, inspired by the range analysis technique, we propose an approach to have a more accurate estimate of the scalar value range over any arbitrary regions. 
The major difference between the proposed technique and the original range analysis technique \cite{sharp2022spelunking} is that we do not produce a ``hard'' bound on the INR output, which tends to be too conservative on a complex scalar field. 
Instead, we treat the input region as an uncertainty source to the INRs and propagate this uncertainty through the neural network to produce an estimated output value probability distribution. This uncertainty propagation is computationally much more efficient than densely reconstructing values inside the region.
The produced output value probability distribution combined with a confidence interval can be used to provide a ``soft'' bound of the output range. 
Using the ``soft'' bounds, we can query the values adaptively in the spatial domain for efficient iso-surface extraction. 
\clrb{Although we are still using an affine approximation to the INR, the ``soft'' bounds are tighter than the range analysis bounds, because we do not consider the extreme cases where all the approximation errors are maximized.}
Although this paper mainly focuses on uncertainty propagation applied to hierarchical iso-surface extraction on INRs, the described technique also applies to the general geometric query tasks on different INRs.

In summary, the contribution of this paper is twofold: 

\begin{itemize}
  \item We present a technique using uncertainty propagation to efficiently estimate the implicit neural representation output distribution over an input domain.

  \item We demonstrate that this output distribution can be easily applied to build a hierarchical data structure for efficient iso-surface extraction on implicit-neural-network-represented data without preprocessing and metadata storage.

\end{itemize}

\section{Related Works}
\label{sect:related_works}
Using coordinate-based neural networks to represent field data, also known as implicit neural representation, is recently a heated research topic in the machine learning literature. A comprehensive survey on the applications and the architectures of INR is given by Xie et al. \cite{xie2022neural}. Since this paper focuses on the efficient iso-surface extraction on INR of scalar fields, in this section, we first survey INR application in scientific data analysis and visualization and then discuss the computation challenge of iso-surface extraction from INRs.

\textit{Applications of INRs in scientific visualization.} The first and most apparent application of INR is to reduce large and high-resolution scalar fields. Lu et al.\cite{lu2021compressive} apply INR and network weight quantization techniques to compress 3D spatial data. Their experiments show that INR-based compression outperforms the traditional state-of-the-art algorithm at the cost of higher encoding and decoding latency. A follow-up study \cite{weiss2022fast} reduces this latency by loading the models to modern GPU tensor cores' on-chip memory and achieves interactive rendering from INRs.
Another work by Pan et al. \cite{pan2021neural} leverages INR as a mesh-agnostic dimensionality reduction tool to generate latent representations for spatio-temporal data.
Despite the direct usage of INR on existing scientific data, other domains like medical imaging and physical simulations also adopt INRs. For example, INRs are used to predict density fields from medical scans of limited viewing angles and sparse views, where the prediction is supervised by mapping the output density value to the sensor domain via Fourier transform (MRI) or Radon transform (CT)\cite{zang2021intratomo, shen2022nerp}. Raissi et al. \cite{raissi2019physics} put forth physics-informed neural networks, where INRs are used to parameterize the solution space of partial differential equations (PDEs) to reduce the simulation computation. Later, this idea is extended to different PDEs \cite{zehnder2021ntopo, izzo2022geodesy, smith2020eikonet, pfrommer2021contactnets}. The wide usage of INRs in these domains motivates an urgent need for efficient visualization techniques for field data represented by neural networks.

\textit{Iso-surface extraction from INRs.} Iso-surface extraction is a well-studied problem in the scientific visualization domain. Many techniques are proposed to improve the iso-surface extraction efficiency. A survey \cite{newman2006survey} on the iso-surface extraction literature summarizes these speed-up algorithms into three categories: hierarchical geometric \cite{wilhelms1992octrees, sutton1999isosurface}, interval-based \cite{livnat1996near, shen1996isosurfacing, cignoni1996optimal} and propagation-based \cite{itoh1995automatic} methods. These speed-up algorithms avoid the computation of non-active cells at the cost of preprocessing and storing metadata. However, when the scalar field is represented as INR, many issues prevent us from applying these techniques. 
First, interval-based and propagation-based techniques cannot be directly used because the INRs describe a continuous field without a structured grid or unstructured mesh, which interval-based methods rely on. Although we can reconstruct values and build the grid from INR, this defeats the purpose of using INR as a succinct and continuous data representation. 
Second, hierarchical geometric methods require the knowledge of the extrema of the values over a region to skip non-active regions. Traditionally, this information is acquired by preprocessing every grid and stored as metadata. However, getting this metadata from INR through dense reconstruction is computationally expensive, takes lots of storage, and is potentially inaccurate if we do not reconstruct dense enough. Sharp and Jacobson \cite{sharp2022spelunking} proposed to apply range analysis or more specifically affine arithmetic, on INRs and bounds the network output over a spatial region directly without dense reconstruction. Their paper not only focuses on iso-surface extraction but proposes a general technique to perform geometric queries on INRs of 3D geometric models. However, due to \clrb{more high-frequency features in} scientific data scalar fields than signed distance functions/occupancy functions for 3D geometric models, range analysis does not give accurate enough bounds in our application and thus leads to poor efficiency. Our method, similar to the range analysis of INR, is focused on but not limited to the application of iso-surface extraction and can be generalized to other visualization tasks that require data queries from INRs. 

In the remainder of this paper, we will explain our technique in detail and present the experiments for hierarchical iso-surface extraction on INRs fitted to various datasets.

\section{Background}
\label{sect:background}

The algorithm proposed in this paper is closely related to the general query technique on INR \cite{sharp2022spelunking}. Range analysis by affine arithmetic is performed on INRs to bound the output over an input region. The output bounds are used to increase the efficiency for different geometric query tasks including hierarchical mesh extraction on signed distance fields, which is very similar to iso-surface extraction from a scalar field.
Before explaining the details of our method, we first introduce the range analysis approach and how it is applied to INRs in this section.

\subsection{Range Analysis via Affine Arithmetic}
\label{sect:ra}
Range analysis is a general technique to bound the function output range over a given input range.
\textit{Interval arithmetic} and \textit{affine arithmetic} are two common methods for range analysis. 
Affine arithmetic decomposes a value interval into the affine combination of different unit intervals. This affine combination is called an affine form. Tracking these unit intervals by applying arithmetic rules guarantees exact output bounds through linear functions.
Formally, an interval $\hat{x}$ is expanded as follows:
\begin{equation}
  \label{eq:affine_form}
  \hat{x} = x_0 + \sum_{i=1}^{N}x_i\epsilon_i, \;\;\;\;\;\;\;
  \epsilon_i \in [-1, 1],
\end{equation}
where each $\epsilon_i \in [-1, 1]$ is a unit interval symbol. 
The scalar value $x_0$ denotes the interval center while each scalar value $x_i$ is the coefficient to scale each unit interval symbol $\epsilon_i$. If either the function input or the output is multidimensional, we can easily extend this definition to intervals in each dimension. 
Intuitively, for any affine form represented interval, we can calculate the original interval when each unit interval takes its minimum and maximum value:

\begin{equation}
  \label{eq:interval}
  \text{range}(\hat{x}) = [x_0-r,x_0+r], \;\;\;\;\;\;\;
  r= \sum_{i=1}^{N}\lvert x_i \rvert.
\end{equation}

\begin{figure}[htbp]
  \centering 
  \includegraphics[width=\columnwidth]{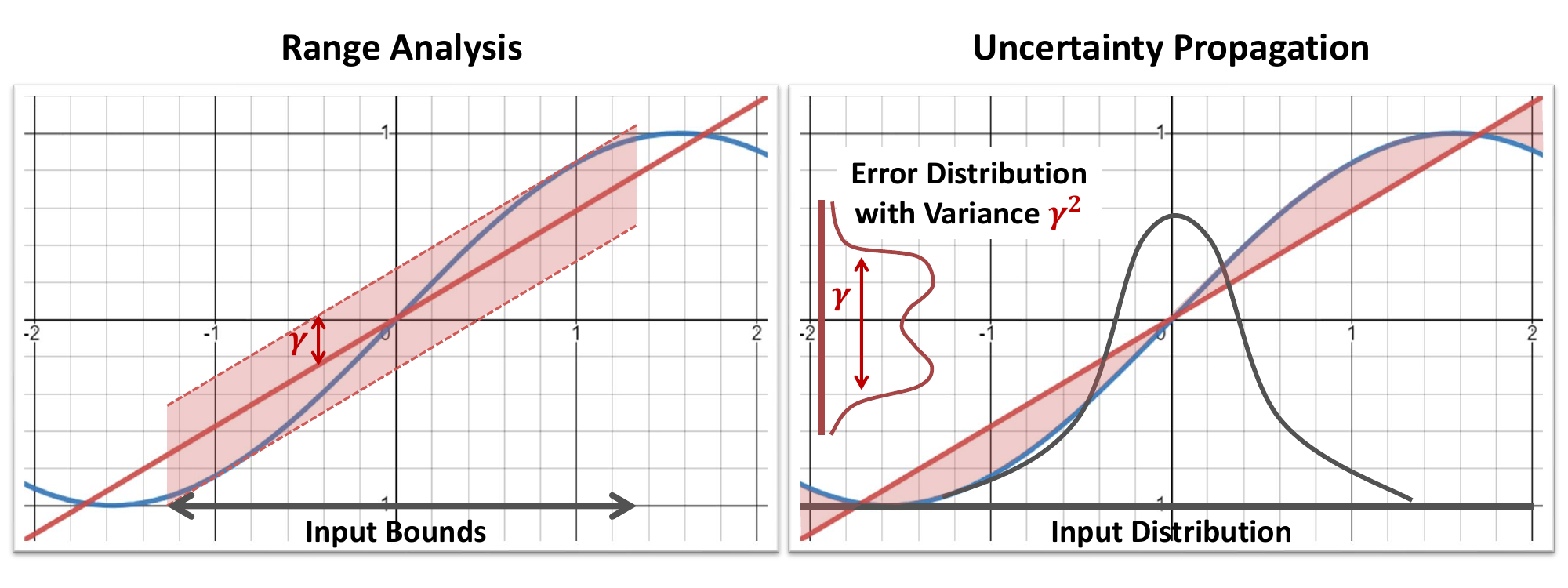}
  \caption{%
  	The difference between range analysis (left) and uncertainty propagation (right) when applying nonlinear functions. The nonlinear function is shown in blue and the linear approximation is in red. \clrb{In the figure on the right, the shaded red region is the error of approximation. Given the input Gaussian distribution, we can calculate the variance $\gamma^2$ of the error distribution. Note that this error distribution is not necessarily Gaussian.}
  }
  \label{fig:difference}
\end{figure}

A nice property of affine forms is that we can directly substitute the input variable in a linear function of affine form and get the exact bound of the function output. In that case, we use the following affine arithmetic rules for a linear function:
\begin{equation}
  \label{eq:linear_rule}
   f(\hat{x}) = \alpha x_0 + \beta + \sum_{i=1}^{N} \alpha x_i \epsilon_i.
\end{equation}
To deal with nonlinear functions, we first obtain a linear approximation to the original function $f(x)\approx\Bar{f}(x):= \alpha x + \beta$. The maximum error between the nonlinear function and the linear approximation is $\gamma=\max_{x\in\text{range}(\hat{x})}{\lvert f(x)-\Bar{f}(x)\rvert}$. 
\clrb{Chebyshev polynomials \cite{mudde2017chebyshev} are used to approximate the nonlinear functions to minimize the maximum error $\gamma$.}
Therefore, we introduce a new unit interval symbol $\epsilon_{N+1} \in [-1,1]$ and the approximation error is bounded by $\gamma \epsilon_{N+1}$. 
Having the error bounded, we can represent the nonlinear function by the affine form in the following way:
\begin{equation}
  \label{eq:nonlinear_rule}
  f(\hat{x}) = \Bar{f}(\hat{x}) + \gamma \epsilon_{N+1} = \alpha x_0 + \beta + \sum_{i=1}^{N} \alpha x_i \epsilon_i  + \gamma \epsilon_{N+1}.
\end{equation}
An example of this approximation can be found on the left of \cref{fig:difference}.

Linear and nonlinear functions can be applied repeatedly to the affine forms (representing the input interval and the intermediate intervals), and the output range can be directly read off from the output affine forms. However, these output bounds from the affine forms are looser than the real output intervals because of the new uncertainty terms $\epsilon_{N+1}$ introduced in the approximation to the nonlinear function.


\subsection{Applying Range Analysis to INRs}
\label{sect:ra_on_INR}
\begin{figure}[htbp]
  \centering 
  \includegraphics[width=\columnwidth]{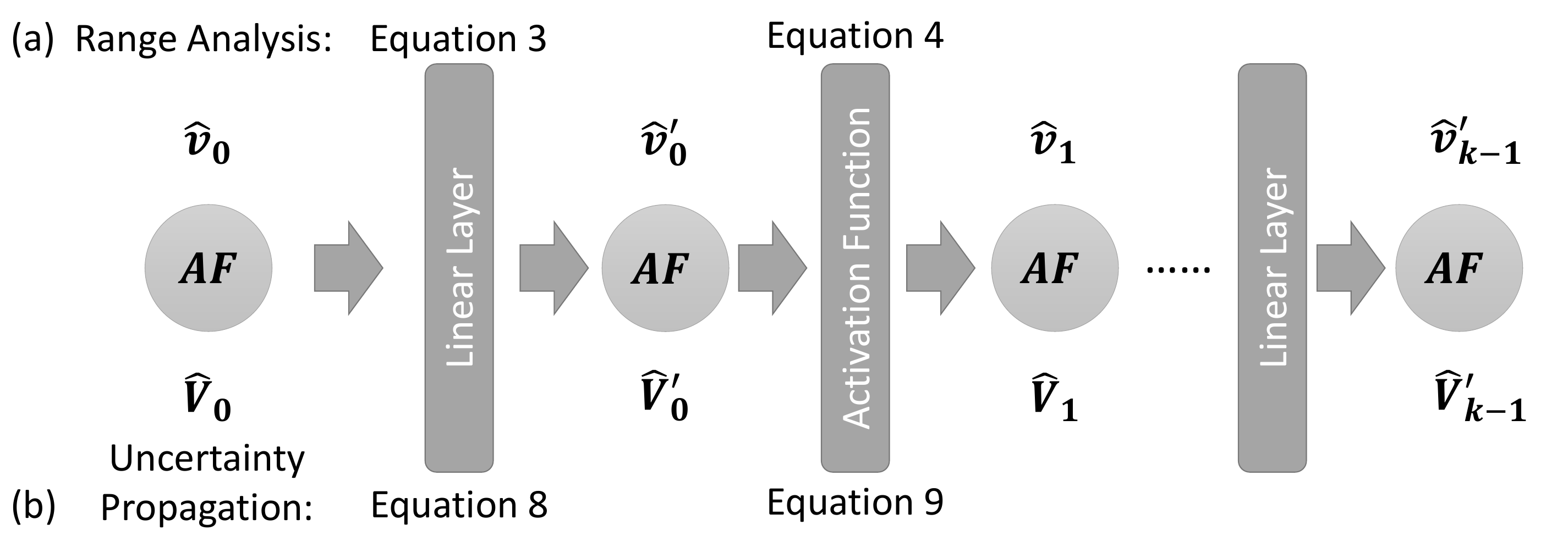}
  \caption{%
  	\clrb{The procedure of applying range analysis and uncertainty propagation through INRs. The scalar values in the network are represented by the affine forms (AF). Range analysis follows \cref{eq:linear_rule} and \cref{eq:nonlinear_rule} while uncertainty propagation follows \cref{eq:paf_linear_rule} and \cref{eq:paf_nonlinear_rule} when applied to INRs.}
  }
  \label{fig:mlp_combined}
\end{figure}

In this section, we explain how the range analysis via affine arithmetic is applied to INRs. 
If we open the black box of neural networks, they are basically a sequence of functions applied to the network input and INR is not an exception. The input to INR is a coordinate tuple describing a spatial position. This input goes through multiple linear layers and nonlinear functions and then maps to an output of the scalar field value at that position. The linear layers in the INR are essentially performing matrix multiplications and the nonlinear function is chosen from activation functions, such as ReLU, ELU, and the sine function. 
If we define an input range to the INR, which could be a 2D or 3D region, performing range analysis via affine arithmetic through INR layers as shown \clrb{in \cref{fig:mlp_combined} (a)}, we can obtain the estimated output range over the region.

The first step is to rewrite the input region in the affine form, and then we can use the rules introduced in \cref{sect:ra} to propagate this affine form through the INR.
Rewriting the input region in the affine form is an intuitive process, for example, a 2D rectangle centered at $x_0$ and $y_0$ with width $2 x_1$ and length $2 y_1$ can be written as an affine form vector $(\hat{x},\, \hat{y})$, where $\hat{x} = x_0 + x_1 \epsilon_1$ and $\hat{y} = y_0 + y_1 \epsilon_2$, with $\epsilon_1, \epsilon_2 \in [-1,1]$. By applying affine transformations to the vector $(\hat{x},\, \hat{y})$, we can represent any parallelogram in the space. A similar approach can be extended to 3D and represent an input region of the rectangular cuboid.

Once the input region is represented in the affine form as the vector $\hat{\bm{v}}_0$ \clrb{in \cref{fig:mlp_combined} (a)}, we can apply the rules in \cref{eq:linear_rule} and \cref{eq:nonlinear_rule} to propagate this range through the network. 
Finally, the estimated scalar value bound can be easily read off from the output affine form $\hat{\bm{v}}'_{k-1}$. These estimated bounds can be used in the hierarchical iso-surface extraction to skip the regions where the iso-value is outside the bound. The details of hierarchical iso-surface extraction are discussed under our approach in \cref{sect:hierarchical_iso_surface}.

\section{Method}
\label{sect:method}
When the range analysis is applied to INRs (discussed in \cref{sect:ra_on_INR}), the number of unit interval symbols $N$ is the sum of the input degree of freedom and the number of activation functions in the network, where a new symbol is introduced when approximating every activation function.
In affine arithmetic, to ensure the query result is always correct, conservative bounds that assume maximum approximation errors are used. However, this is unlikely to happen in practice. As mentioned in the introduction, we observe that the range analysis bounds are too conservative, resulting in poor efficiency when used for hierarchical iso-surface extraction.

Based on this insight, we propose our approach that estimates the output value distribution over an input region to the INR instead of only the output bounds. Combining the estimated distribution and a confidence interval threshold, we can generate a ``soft'' bound for the INR output. To extract the iso-surfaces, we build a k-d tree in a top-down manner. The ``soft'' bounds are used to determine whether or not the region corresponding to a k-d tree node contains a component of the iso-surface. If yes, the space is subdivided adaptively and we only reconstruct the values densely at the regions of interest. 

To estimate the scalar value distribution, we first propose a straightforward method assuming each unit interval symbol from range analysis represents a uniformly distributed random variable in \cref{sect:RA-UA}. With that assumption, we can easily estimate the distribution from the range analysis output. However, this simple technique does not generate an accurate enough estimate. To solve this problem, in \cref{sect:paf} we propose to use another technique called probabilistic affine forms (PAFs), which models the input to functions as random variables that are affine combinations of unit random variables. By tracking these PAFs through functions, we can get PAFs as the function output. We estimate these PAFs' distributions as Gaussian distributions by applying the central limit theorem (CLT). We expect this approach to generate a more accurate distribution estimation compared to the range analysis with uniform assumptions. The process that tracks PAF through functions is called uncertainty propagation.
In the rest of this section, we provide the details of the technique and explain how it is used in hierarchical iso-surface extraction.

\subsection{Range Analysis with Uniform Assumption}
\label{sect:RA-UA}
The affine form expressed in \cref{eq:affine_form} is merely a combination of the unit interval symbols, which does not tell the value probability distribution inside the interval. We first propose a straightforward approach that assumes the values are uniformly distributed inside each unit interval. To put it formally, for an affine form $\hat{x} = x_0 + \sum_{i=1}^{N}x_i\epsilon_i$, we assume each $x_i\epsilon_i$ follows a uniform distribution defined between $[-x_i, x_i]$. To calculate the exact probability distribution of $\hat{x}$, we need to convolve the probability density function of each $x_i\epsilon_i$, which is too computationally expensive when $N$ is large. Fortunately, we can estimate the probability distribution of $\hat{x}$ as a Gaussian distribution using the central limit theorem (CLT). According to the CLT, the sample averages drawn from a sequence of $n$ identically and independently distributed random samples converge in distribution to Gaussian as $n \rightarrow \infty$. This also indicates the sums of the samples converge in distribution to Gaussian. However, since in the affine forms, each $x_i\epsilon_i$ is not identically distributed, we resolve to a specific variant of the CLT that relaxes the condition to non-identically distributed random variables called Lindeberg CLT\cite{lindeberg1922neue}.
The theorem states that for a sequence of independent random variables $\{X_1, ..., X_n, ...\}$, each with the mean $\mu_i$ and the variance $\sigma_i^2$, the sum of $(X_i-\mu_i)$ converges in distribution to a Gaussian random variable as $n\rightarrow\infty$:
\begin{equation}
  \label{eq:Lindeberg_clt}
  \sum_{i=1}^{n}{(X_i-\mu_i)} \xrightarrow[]{d} \mathcal{N}(0,s_n^2), \;\;\;\;\;\;\; s_n^2 = \sum^n_{i} \sigma_i^2.
\end{equation}
\clrb{Lindeberg CLT also requires that the random variables satisfy Lindeberg's condition\cite{lindeberg1922neue}. This condition enforces that no random variable has a variance that is significantly larger than others. }

For a uniformly distributed random variable, variance is calculated as $\sigma^2 = \frac{1}{12}(x_\text{\clrb{upper}} - x_\text{lower})^2$.
Therefore each symbol $x_i \epsilon_i$ ($\epsilon_i \in [-1, 1]$), under the uniform distribution assumption, has $\mu_i=0$ and $\sigma_i^2 = \frac{1}{3}x_i^2$. Applying \cref{eq:Lindeberg_clt},
the summation $\sum_{i=1}^{N}x_i\epsilon_i$ follows a Gaussian distribution $\mathcal{N}(0, s_N^2)$, where $s_N^2 = \frac{1}{3} \sum_{i=1} ^ N  x_i^2$. Therefore, we estimate the affine form $\hat{x}=x_0 +\sum_{i=1}^{N}x_i\epsilon_i$ to follow a Gaussian distribution $\mathcal{N}(x_0, s_N^2)$. 
\clrb{The independence, Lindeberg's condition, and the large number assumption cannot be strictly proved for our application on INRs, since no restriction is placed on the INR weights. However, we find the assumptions heuristically hold in our experiments and discuss the situations when these assumptions might break in \cref{sect:discussion}.}
Having this assumption, we can easily estimate the INR output distribution over an input region from the range analysis result and this distribution can be used to generate the ``soft'' bounds for iso-surface extraction, which we discuss in \cref{sect:hierarchical_iso_surface}. We call this simple modification range analysis with the uniform assumption (RA-UA)


However, there are two pitfalls in RA-UA. The first is that the uniform assumption of error may not hold on different activation functions in INR. Although the CLT can be applied to any distribution shape, the variance calculated is incorrect if $\epsilon_i$ does not follow a uniform distribution. 
The second issue is that the distribution estimation only happens at the final INR output. In the intermediate layers, the ``hard'' bounds are still used to calculate $\gamma$ in \cref{eq:nonlinear_rule}. This already introduces intervals that contain values with low probability, which will eventually lead to pessimistic results especially when the INR network is deep.

\subsection{Uncertainty Propagation}
\label{sect:paf}
Because of the two downsides mentioned in the last section, we further propose uncertainty propagation through INR using the probabilistic affine form (PAF), where the probability estimation is performed in each layer of activation function approximation and, the exact variance of each estimation error can be calculated thanks to the property of Gaussian distribution. Uncertainty propagation is expected to have a more accurate output distribution estimate compared to the RA-UA which we will use as the baseline method in the evaluation.

Probabilistic affine forms for uncertainty propagation were previously proposed by Bouissou et al. \cite{bouissou2012generalization, bouissou2016uncertainty}. 
These two papers describe a more general PAF considering arbitrary distributions and arbitrary dependency between random variables. However, for computation efficiency, we only consider a special case where random variables are independent and we approximate their linear combination with Gaussian distributions in this study. Therefore, the notations and definitions used in this paper are different from the previous papers for a simpler and clearer explanation. We discuss our PAF in this section and recommend the readers who are interested in a general and complete PAF definition to the previous papers.


First, we formally define a random variable $\hat{X}$ as an affine combination of multiple unit random variables $Z_i$:
\begin{equation}
  \label{eq:paf}
  \hat{X} = x_0 + \sum_{i=1}^{N}x_iZ_i.
\end{equation}
Each $Z_i$ can have an arbitrary probability distribution, however, we enforce each $Z_i$ has unbiased mean and unit standard deviation ($\mathbb{E}(Z_i)=0$ and $\text{Var}(Z_{i})=1$). We will discuss how this is enforced later in this section.
We also assume unit random variables to be independent of each other, \clrb{so that the affine form can be approximated with a Gaussian distribution using the CLT.}

The probability density function (PDF) of $\hat{X}$ is the convolution of each individual PDF of $Z_i$. For $Z_i$ that is arbitrarily distributed and does not have a general closed-form PDF expression, convolving the PDF of a large number of $Z_i$ is too computationally expensive to fit our application. Therefore, similar to the range analysis with uniform assumption discussed in \cref{sect:RA-UA}, we decide to approximate the distribution of $\hat{X}$ with Gaussian distributions using Lindeberg CLT. 
\clrb{Because in PAFs, the mean of each term $x_i Z_i$ is zero and the variance is $x_i^2$.}
Following the \cref{eq:Lindeberg_clt}, the distribution of $\hat{X}$ can be estimated as:
\begin{equation}
  \label{eq:clt}
  \hat{X} \sim \mathcal{N}(x_0,\,s_N^2), \;\;\;\;\;\;\; s_N^2=\sum_{i=1} ^ N x_i^2.
\end{equation}
Similar to RA-UA, the assumptions that enable us to apply Lindeberg CLT are discussed in \cref{sect:discussion}.

The PAF can be similarly propagated through functions like affine forms in the range analysis. The rule to apply linear functions to PAF is similar to affine arithmetic as described in \cref{eq:linear_rule}:
\begin{equation}
  \label{eq:paf_linear_rule}
  f(\hat{X}) = \alpha x_0 + \beta + \sum_{i=1}^{N} \alpha x_i Z_i,
\end{equation}
where $\alpha$ and $\beta$ are the coefficients for the linear function. The random variable distribution is exact 
through linear functions because of the property of affine arithmetic. It is worth noting that we may have errors when modeling the distribution of $\hat{X}$ with Gaussian distribution following Lindeberg CLT. However, this inaccuracy is not introduced by the linear function.

More differences exist between applying nonlinear functions to PAFs and affine forms. The nonlinear function is first approximated with a linear function $f(x)\approx\Bar{f}(x):= \alpha x + \beta$ and we introduce a new random variable $\gamma Z_{N+1}$ that denotes the error of this approximation. 
\clrb{In the right of \cref{fig:difference}, we use the shaded red region to denote the approximation error.}
Then the application of the nonlinear function is converted to a linear function plus the approximation error:
\begin{equation}
  \label{eq:paf_nonlinear_rule}
  f(\hat{X}) = \Bar{f}(\hat{X}) + \gamma Z_{N+1} = \alpha x_0 + \beta + \sum_{i=1}^{N} \alpha x_i Z_i + \gamma Z_{N+1}.
\end{equation}
Because we want to enforce that the random variables in the PAF have unit standard deviations, $\gamma$ is set to be the standard deviation of the approximation error so that $\text{Var}(Z_{N+1})=1$.
\clrb{The approximation error $\gamma Z_{N+1}$ may have an arbitrary PDF determined by the activation function, approximation method, and the input distribution $\hat{X}$. However, because we approximate the distribution of the PAF using the CLT as shown in \cref{eq:clt}, calculating the error variance $\gamma ^2$ is sufficient to propagate the PAF through a nonlinear function. }

The next step is to choose a method for the linear approximation. We choose the least square method for the following two reasons. First, the least square method minimizes the mean squared error (MSE) and the solution is unbiased. This ensures the introduced approximation error bias $\mathbb{E}(Z_{N+1}) = 0$ and minimizing the MSE is the same as minimizing the error variance $\gamma^2$,
because the MSE can be decomposed into the sum of the error variance and the squared bias: $\text{MSE} = \gamma^2 + [\mathbb{E}(Z_{N+1})]^2$.
Second, the linear least square method has the analytical solution to approximating different activation functions, which allows us to perform the propagation efficiently. To put it formally, given that the nonlinear function input PAF $\hat{X}$ follows a Gaussian distribution $\mathcal{N}(\mu, \sigma^2)$ with PDF $g(x |\, \mu, \sigma^2)$, the MSE is an integral of the squared error overs this Gaussian PDF. Therefore, the optimization of the MSE is defined as:
\begin{equation}
    \label{eq:least_square}
    \argmin_{\alpha,\, \beta}{\int_{-\infty}^{\infty} g(x |\, \mu, \sigma^2)(f(x)-\alpha x - \beta)^2\dif{x}},
\end{equation}
where $\alpha, \beta$ are parameters for the linear approximation.
This minimum can be found by solving the following system:
$\frac{\partial \text{MSE}}{\partial\alpha} = 0, \,
\frac{\partial \text{MSE}}{\partial\beta} = 0$. 
\clrb{The analytical solution for the three commonly used activation functions, ReLU, ELU, and Sine, can be found in the appendix.}


\subsection{Uncertainty Propagation through INRs}
\label{sect:up_INR}
Similar to the range analysis, uncertainty propagation through INRs also sequentially applies linear and activation functions on PAFs. We summarize these processes \clrb{in \cref{fig:mlp_combined} (b)}. The input region is represented as a vector $\hat{V}_0$ of PAFs. After the uncertainty propagation through an INR with $k$ layers, we get an output PAF $\hat{V}'_{k-1}$.


The input region $\hat{V}_0$ is represented by an affine combination of uniformly distributed random variables. 
For example, to describe 2D square centered at $x_0$ and $y_0$ with width of $2 x_1$ and length of $2 y_1$, the input PAF vector is $(\hat{X},\, \hat{Y})$, where $ \hat{X} = x_0 + \frac{x_1}{\sqrt{3}} Z_1$ and $ \hat{Y} = y_0 + \frac{y_1}{\sqrt{3}} Z_2$. The coefficients $\frac{x_1}{\sqrt{3}}$ and $\frac{y_1}{\sqrt{3}}$ normalize the uniform random variable $Z_1$ and $Z_2$ to have unit standard deviations since a uniform distribution defined on $[-x_1,x_1]$ has standard deviation $\sigma = \frac{x_1}{\sqrt{3}}$. We can define rectangles with arbitrary shapes at any position by applying an affine transformation to $(\hat{X},\, \hat{Y})$. A similar input definition can also be extended to 3D.

Once the input region is written as a PAF vector, we can propagate it through the INR linear layers and activation functions using the rules defined by \cref{eq:paf_linear_rule} and \cref{eq:paf_nonlinear_rule}. 
In the last layer, we get the INR output distribution over the input region from the PAF. The output distribution is used to generate ``soft'' bounds for the hierarchical iso-surface extraction, which we will introduce in the next section.

\subsection{Hierarchical Iso-surface Extraction}
\label{sect:hierarchical_iso_surface}

A trivial and straightforward way to extract iso-surface from the INR is through dense reconstruction. One can decide the dense grid resolution based on the maximum error that can be tolerated. After that, the marching cubes algorithm can be applied inside the grid assuming scalar values in the cells are linearly interpolated. 
To avoid densely reconstructing the data from INR, in this section, we propose a hierarchical iso-surface extraction algorithm using the k-d tree. We estimate the value distribution inside the tree nodes through the technique discussed in \cref{sect:up_INR} and use its confidence interval to adaptively build this tree with respect to the chosen iso-value and identify the potential active cells. 
The data reconstruction from the INR is only performed on the active cell corners to reduce computation. We present the pseudocode for the active cell finding in \cref{alg:extraction}. 

\begin{algorithm}
	\caption{KdTreeExtraction($f_\theta, \bm{x}_l, \bm{x}_u, c, t$)} 
    \label{alg:extraction}
	\begin{algorithmic}[1]
    \Require {An INR $f_\theta: \mathbb{R}^3 \rightarrow \mathbb{R} $ of the scalar field, domain bounds $\bm{x}_l,\,\bm{x}_u \in \mathbb{R}^3$, iso-value $c$, and a hyperparameter $t$.}
    \Ensure {Predicted set $A$ of the active cells.}
    \State $\bm{x}_c \gets (\bm{x}_l + \bm{x}_u) /2$
    \Comment{Node Center}
    \State $V \gets \text{Diag}{(\frac{1}{\sqrt{3}}(\bm{x}_u-\bm{x}_c))}$ \Comment{Diagonal matrix for orthogonal axes}
    \State $\hat{\bm{X}} \gets \bm{x}_c + V \bm{\epsilon}$
    \Comment{Vector of PAFs in matrix notation }
    \State $\hat{\bm{Y}} \gets \text{UP}(f_\theta, \hat{\bm{X}})$
    \Comment{Uncertainty propagation in \cref{sect:up_INR}}
    \State $\mu, \sigma \gets \text{CLT}(\hat{\bm{Y}})$
    \Comment{Get $\mu$ and $\sigma$ from output PAF using \cref{eq:clt}}
    \If{$\mu + t \sigma < c $ \Or $ \mu - t \sigma > c$}
    \State \Return $\emptyset$
    \Else
    \If {Max depth reached}
    \State \Return $\{(\bm{x}_l, \bm{x}_u)\}$
    \Else
    \State $\bm{x}_u^{'}, \bm{x}_l^{'} \gets \text{Split}(\bm{x}_u, \bm{x}_l)$
    \Comment{Split along the widest dimension}
    \State $A_{\text{left}} \gets \text{KdTreeExtraction}(f_\theta,\bm{x}_l, \bm{x}_u^{'}, c, t)$
    \State $A_{\text{right}} \gets \text{KdTreeExtraction}(f_\theta,\bm{x}_l^{'}, \bm{x}_u, c, t)$
    \State \Return $A_{\text{left}} \cup A_{\text{right}} $
    \EndIf
    \EndIf
	\end{algorithmic} 
\end{algorithm}

The k-d tree \cite{bentley1975multidimensional} is built from top to bottom to subdivide the domain and find the finest active cells that contain a component of the iso-surfaces. 
The finest active cell is determined by the maximum depth of subdivision which is related to the computation resources we have.
Starting from the root node, we describe the corresponding 3D region as a bounding box with lower and upper corners $\bm{x}_l$ and $\bm{x}_u$. To represent this region as a vector of PAFs, the center and the orthogonal axes of the box are calculated (see steps one through three in \cref{alg:extraction}). 
Having the PAF vector, we propagate it through the INR and get the output PAF $\hat{\bm{Y}}$. \clrb{Applying the CLT to the output PAF, we estimate the distribution of scalar values inside the node region with a Gaussian distribution with mean $\mu$ and standard deviation $\sigma$.}
A confidence level value $t$ bounds the confidence interval of this region to be $[\mu - t \sigma, \mu + t \sigma]$. 
A larger $t$ will give more conservative bounds and a smaller $t$ will give tighter bounds. Detailed studies on the choices of hyper-parameter $t$ can be found in \cref{sect:hyper}.
The node is split when the interested iso-value is inside the bounds. Otherwise, the node is marked as non-active and skipped for further subdivision.
We always split along the widest dimension in our algorithm, which ensures all three dimensions will be subdivided in turns. 
After subdivision, the algorithm is applied to the child nodes recursively. 
When the iso-value is inside the estimated bound AND the maximum depth of the k-d tree is reached, we mark this leaf node as an active cell. 
We enforce the maximum depth of the tree to be a multiple of three so that the active cell is always a cube.
All active cells are collected as the algorithm output.
After collecting all the predicted active cells, the grid corners of the active cells can be found.
We query the INR for the scalar value at the grid corners and perform the marching cubes algorithm to extract the triangle meshes inside the cells.

\section{Experiments}
\label{sect:experiments}

We perform experiments to evaluate our proposed method from different perspectives. First, the hierarchical iso-surface extraction results from 
different datasets are reported to show the efficiency and the quality of the approach. 
Second, we extract the iso-surfaces in higher resolutions (larger k-d tree max depth) to test the scalability of the approach.
Third, we directly compare our estimated value distribution to the true value distribution over a region calculated by Monte Carlo sampling to show the accuracy of our estimation.

\subsection{Experiment Settings}
In this section, we discuss the experiment setting including datasets used for experiments, the INRs for the datasets, the evaluation metrics, and the compared baseline methods.
\subsubsection{Datasets}
We evaluate our iso-surface extraction efficiency and accuracy on four scalar fields from different datasets. 
Datasets are chosen to cover different resolutions and different densities of the iso-surfaces in the domain.
We first briefly introduce the datasets used. 
\textbf{The vortex dataset (Vortex)} is from a pseudo-spectral turbulent simulation performed by the mechanical engineering department at Rutgers in 1995. The simulation outputs the vorticity magnitude field with spatial resolution $128\times128\times128$ across $30$ time steps. We use one specific time step which clearly shows the vortex features for the experiment.
\textbf{The ethanediol dataset (Ethanediol)} is obtained from the topology toolkit \cite{tierny2017topology}. The dataset contains bi-variate scalar fields of resolution $115 \times 116 \times 134$ and describes an ethanediol molecule. \clrb{We use one of the scalar fields for the experiments.}
\textbf{The turbulent combustion dataset (Combustion)} is the output from a direct numerical solver called S3D produced by Sandia National Laboratories. The available data contains a vorticity magnitude field of dimension $480 \times 720 \times 120$, which is used for our experiments.
\textbf{The force isotropic dataset (Isotropic)} from Johns Hopkins Turbulence Databases \cite{kanov2015johns} is the result of a direct numerical simulation solving Navier-Stokes equations using the pseudo-spectral method. The dataset is in $1024^3$ resolution and we calculate the vorticity magnitude field for our experiments.

\subsubsection{INR models}
\label{sect:INR_models}
We use a SIREN \cite{sitzmann2020implicit} based architecture to train the INRs on different datasets, for its good representation power with a relatively small number of parameters. 
The SIREN is composed of multiple linear layers and sinusoidal activation functions. For the vortex, the ethanediol, and the combustion dataset, we use a relatively small network (8-layer and 32-width). For the isotropic dataset, we use a larger SIREN with three 128-width layers. 
A study on the network size's influence on extraction time and accuracy is performed in our evaluation.
\clrb{We also present the experiment results on INRs with ReLU and ELU activation functions in the appendix.}
INR representation size, quality, and the original data size are shown in \cref{tab:psnr}. The reconstruction quality is calculated in peak signal-to-noise ratio (PSNR), which is a widely used metric for image and volumetric data quality.
Our INR fitting quality is similar to the results shown by other studies \cite{lu2021compressive, weiss2022fast}. \clrb{We analyze the INR qualitatively by volume rendering the densely resampled INR in the appendix. The scientific scalar fields contain much more high-frequency features compared to SDFs of 3D models.} 

\begin{table}[htbp]
  \caption{%
  INR reconstruction quality on different datasets.
  }
  \label{tab:psnr}
  \scriptsize%
  \centering%
  \begin{tabu}{%
  	  r%
  	  	*{4}{c}%
  	}
  	\toprule
  Dataset & Vortex & Ethanediol & Combustion & Isotropic   \\
  	\midrule
   Size (MB) & 8.1 & 6.9 & 159 & 5461 \\
   INR Size (KB) & 37 (0.44\%) & 37 (0.52\%) & 37 (0.002\%) & 135 (0.0002\%) \\
  	PSNR (dB) & 38.23 & 54.87 & 37.78 & 26.89 \\
  	\bottomrule
  \end{tabu}%
\end{table}

\subsubsection{Metrics}
\label{sect:metrics}
To evaluate the iso-surface extraction efficiency, we calculate the relative time of one single inference of the range analysis or the uncertainty propagation compared to an ordinary scalar evaluation of the INR. We also report the average volume of the non-active nodes in our experiments. We calculate the volume assuming the total volume of the domain is one.
Larger sizes of the non-active nodes indicate that more nodes are skipped in the lower level of the kd-tree.
Smaller computation time and larger non-active cells both lead to higher iso-surface extraction efficiency.
We also calculate and compare the total iso-surface extraction time, which is the sum of the time to predict active cells (either by range analysis or uncertainty propagation), the time for network inference on active cells, and the time for generating triangle mesh inside the active cells.

\clrb{
There are two types of errors in the hierarchical iso-surface extraction. The first type is false positive, i.e. non-active cells incorrectly predicted as active cells. The second type is false negative, i.e. active cells incorrectly predicted as non-active cells. 
False positives will lead to longer computation time but will not have incorrect extraction results. However, false negatives will lead to erroneous extraction results.
Therefore, we calculate both the false negative rate (FNR) and the false positive rate (FPR) of our proposed method and the baseline methods. FPR is defined as $\frac{FP}{FP+TN}$ and FNR is defined as $\frac{FN}{FN+TP}$, where FN, FN, TP, and TN denote the number of false negative, false negative, true positive, and true negative cells respectively. These two scores are in the range of $[0,1]$. Lower values are better.}

\subsubsection{Baselines}
Our uncertainty propagation approach (\textbf{UP}) is compared to six different baselines for evaluation. The first and the most straightforward baseline is applying marching cubes to the densely reconstructed grid, which is called \textbf{Dense} in the remaining sections. For comparison, the grid resolution is chosen to match the maximum level of the k-d tree in our approach. 
The second baseline is hierarchical iso-surface extraction through range analysis (\textbf{RA}) and its variants. Including the vanilla variant, there are four RA variants, namely, 
\textbf{RA-Full}, \textbf{RA-Fixed}, \textbf{RA-Truncate}, and \textbf{RA-Append}.
RA-Full indicates the vanilla range analysis method introduced in \cref{sect:background}, while the other variants aim to reduce the computation of range analysis by combining the RA coefficients at the expense of the tightness of the bounds.
Details about how RA coefficients are combined in these variants can be found in the original paper \cite{sharp2022spelunking}. 

The third baseline is range analysis with the uniform assumption (\textbf{RA-UA}), a simple modification of the range analysis technique to tighten the output bounds that is already introduced in \cref{sect:ra}.

\subsubsection{Implementation}

Our implementation of the uncertainty propagation follows the JAX framework \cite{jax2018github} used by Sharp and Jacobson \cite{sharp2022spelunking}, which provides an easy and efficient way to combine neural networks with the range analysis or the uncertainty propagation.
The network training, inference, and range estimation are all accelerated with an A100 GPU on a machine with a 16-core 2.3GHz processor.
A GPU-accelerated parallel marching cubes algorithm that is similar to the one in Nvidia’s CUDA SDK is implemented under the JAX framework. 
The source code of the implementation and the corresponding guides to run the code are available on OSF at \href{https://osf.io/dg6rc/?view_only=ec94d6a3454845dc98c7da88d585c0d3}{osf.io (link)}, released under the MIT license.

\subsection{Hyperparameter Choice}
\label{sect:hyper}
The first hyperparameter is the target resolution for iso-surface extraction, which is chosen based on the complexities of the different datasets. This resolution should be the same for all baseline methods and our method.
The target resolution for the vortex and the ethanediol dataset is $256^3$ (max k-d tree depth $d=24$). For the combustion dataset, we extract the surface at the $512^3$ ($d=27$) resolution and the isotropic dataset at the $1024^3$ ($d=30$) resolution.

\begin{figure}[htbp]
  \centering 
  \includegraphics[width=\columnwidth]{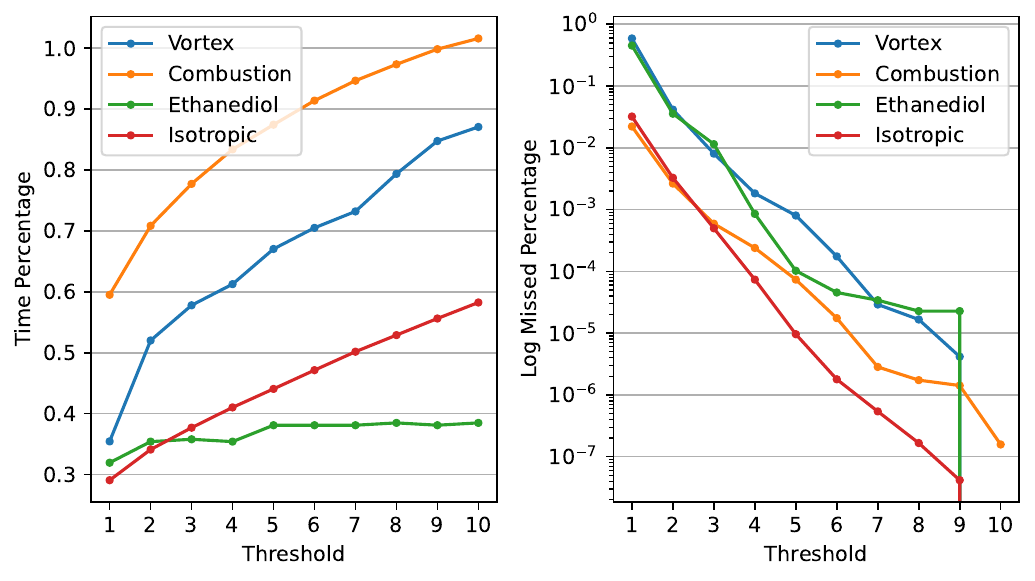}
  \caption{%
  	We compare exaction time and quality under different hyperparameter choices from $t=1$ to $t=10$. The time is calculated as a percentage of the dense reconstruction time and the quality is indexed by the ratio of the missed iso-voxel out of the total number of iso-voxels in the log scale.
  }
  \label{fig:hyper}
\end{figure}

The confidence level value $t$ (introduced in \cref{sect:hierarchical_iso_surface}) is a hyperparameter of our approach, which controls the tightness of the ``soft'' bound. Larger $t$ will make it less likely to mistakenly skip the active cell but increase the time to subdivide the tree and reconstruct more data points. Therefore, we expect the choice to be a trade-off between quality and extraction time.
To analyze the influence of $t$ on the iso-surface extraction time and quality, we perform the experiments using the settings from $t=1$ to $t=10$. The time and quality results are shown for all four datasets in \cref{fig:hyper}.
We observe that the extraction time roughly increases in a linear manner with the increase of $t$, while the number of missed iso-voxels decreases exponentially (the right figure in \cref{fig:hyper} is in the log scale) with the increase of $t$. Therefore, the results indeed confirm that the choice of $t$ is a trade-off.  

\begin{figure}[htbp]
  \centering 
  \includegraphics[width=\columnwidth]{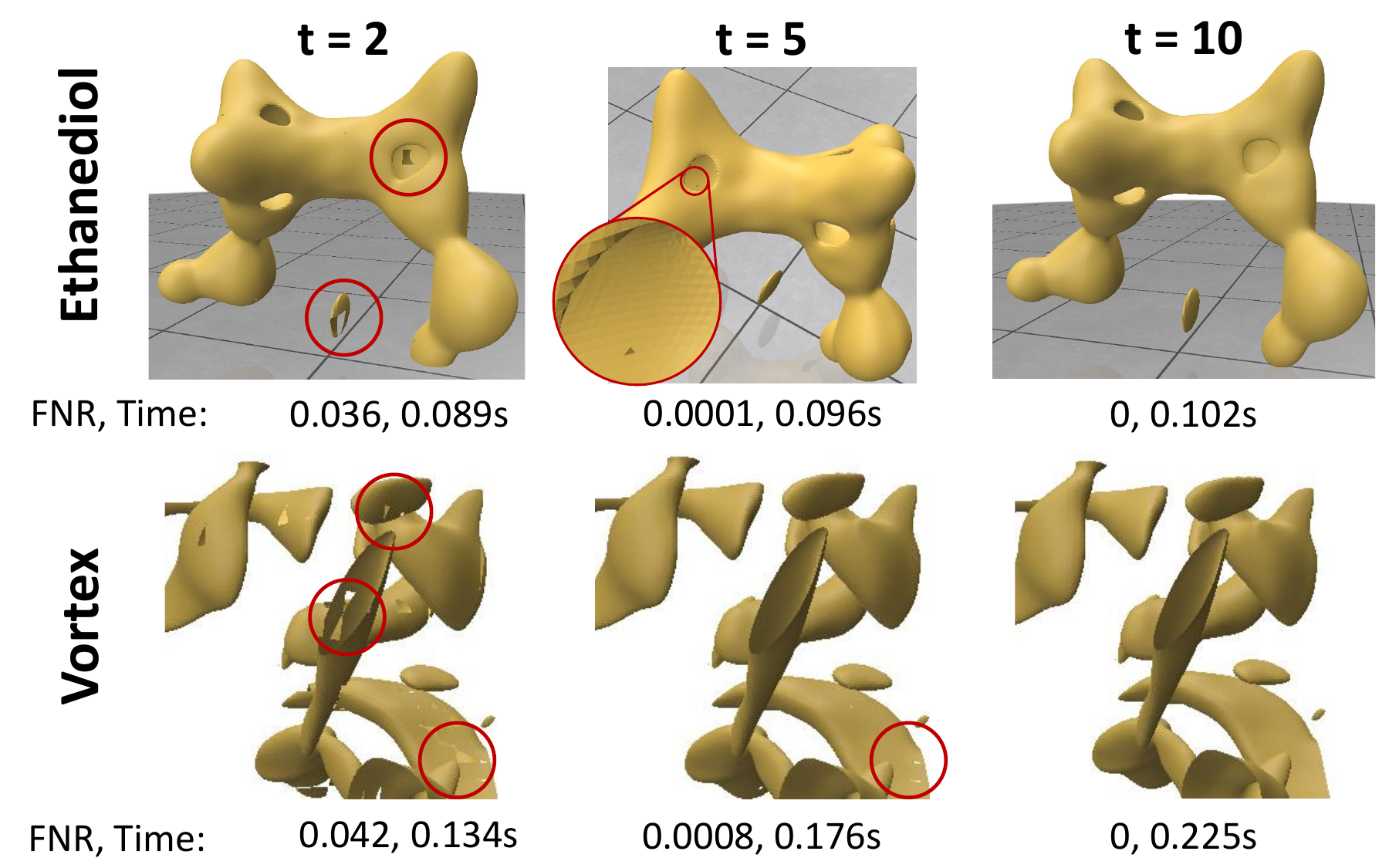}
  \caption{%
  	\clrb{The figure shows the extraction results, false negative rate (FNR), and extraction time of two datasets using different t values.}
  }
  \label{fig:t_qualitative}
\end{figure}

\clrb{We also show some rendered extraction result in \cref{fig:t_qualitative}. When $t=2$, there are many obvious missing components in the extraction. When $t=5$, there are only subtle missing components that can be found with careful examination. When $t=10$, the extraction results have no missing components for these two datasets. We can see that $t=5$ is a good trade-off between computation time and quality. The extraction error is also data-dependent. For example, the vortex dataset has overall higher errors compared to the Ethandiol dataset under all $t$ settings. Therefore, the value of $t$ should be tweaked based on the use case and the dataset. }
It is worth noting that even when $t=10$, the extraction time is still shorter than the time of the baseline methods and the dense reconstruction, and the extraction results are perfect for almost all tested datasets. 
\clrb{In the rest of our experiments, we present the results of our approach with $t=2$, $t=5$, and $t=10$.}

\subsection{Iso-surface Extraction Time}
Before reporting the total time for iso-surface extraction, we first present the relative inference time of different approaches (UP, RAUA, RA, etc.) and the averaged non-active node size.
\clrb{The experiments are performed on the model with 8 layers and 32 neurons in each layer.}
\clrb{Theoratically, the relative inference time will increase when the number of parameters in the network is increased, because the number of uncertainty source $N$ in \cref{eq:paf} is increased. The increase on relative inference time of our method and RA-Full will be in the same order, while other RA variants can be more scalable.}
The relative inference time for uncertainty propagation is $99.8$ times an ordinary scalar evaluation of the network, compared to $126.2$ times for RA-Full, $19.6$ times for RA-fixed, $102.5$ times for RA-Truncate, and $29.1$ times for RA-Append. 
In terms of the time for \textbf{single} inference, our approach is faster than the vanilla RA variant (RA-Full) but slower than other variants. This is expected because other range analysis variants trade range estimation accuracy for computation time. 
However, less accurate estimation will increase the total count of network inference and thus make the total computation time longer. 
To show this, we also calculate the average volume of the non-active nodes for different methods. A larger non-active node indicates less network inference is needed for the active nodes. Assuming the total volume of the domain is one, our approach has an average skipped volume of $7*10^{-4}$, which is about $20$ times the average volume $0.35 * 10^{-4}$ for RA-Full. Other variants' estimation is too loose to skip any non-active node. In that case, we are not able to calculate an average volume. However, not skipping any node in the tree structure shows these variants fall back to dense reconstruction.

\begin{figure}[htbp]
  \centering 
  \includegraphics[width=\columnwidth]{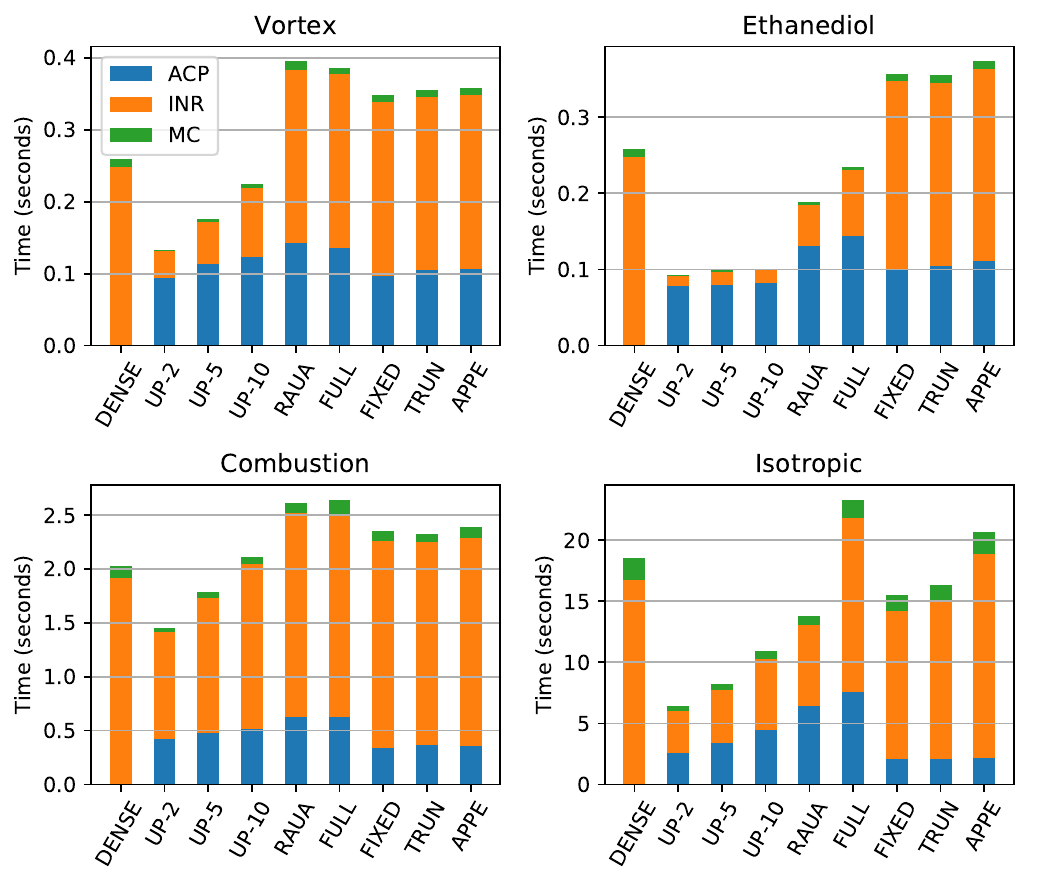}
  \caption{%
    Iso-surface extraction time (in seconds) comparison between our method UP \clrb{with different $t$ (UP-$t$)} and the baseline methods on different datasets. The extraction time is separated into the active cell prediction time (ACP), the data reconstruction time (INR), and the surface generation time (MC). 
  } 
  \label{fig:time}
\end{figure}

We present the total iso-surface extraction time in \cref{fig:time}. 
Among all approaches, ours has the smallest extraction time compared to dense reconstruction and different variants of range analysis.
The vanilla range analysis (Full) works poorly in almost all datasets. The INR reconstruction time is similar to that of the dense reconstruction meaning not many cells are skipped, however, the overhead to performing range analysis makes it consume more computation than the dense reconstruction. 
These observations indicate the RA-generated bounds are too conservative.
Other variants (fixed, truncate, and append) reduce the computation for active cell prediction as shown by the blue bars in \cref{fig:time}. As a result, these variants may be faster than RA-Full in some datasets.
However, the range analysis bounds of these variants are more \clrb{conservative}, leading to more false positive cells and longer INR reconstruction time. 
Ultimately, these RA variants take longer time than our approach in all datasets.
The INR inference time for the other baseline, range analysis with the uniform assumption (RA-UA), is slightly shorter than the dense reconstruction and vanilla range analysis. This shows that RA-UA generates slightly less pessimistic bounds. However, the extraction time is still much longer than our uncertainty propagation approach (UP).

Additionally, across all approaches, the marching cubes computation time is much less compared to the INR inference time. More sophisticated algorithms like the one proposed by Liu et al. \cite{liu2016parallel} can be used to further decrease the marching cubes computation time. Nevertheless, replacing the marching cubes algorithm with a different one will have similar effects on our method and the baseline methods. In essence, the marching cubes algorithm used in the experiments is only to demonstrate the iso-surface extraction results. It's independent of our proposed approach and therefore not the focus of this paper.

\subsection{\clrb{Errors in Active Cell Prediction}}

\clrb{Two types of errors, i.e. false positive error and false negative error, could happen in the process of hierarchical iso-surface extraction. False positive errors incorrectly predict non-active cells as active cells leading to prolonged computation time. Inversely, false negative errors predict active cells as non-active leading to missing components in the iso-surface extraction results. We first present the quantitative results for the false positive and false negative in \cref{tab:f_score}. }

\begin{table}[htp]
  \caption{%
  \clrb{FPR ($\frac{FP}{FP+TN}$) and FNR ($\frac{FN}{FN+TP}$) of the active cell prediction using different approaches. Lower FPR and FNR are better. The best score is highlighted in bold text. UP-$t$ represents uncertainty propagation using different thresholds.}
  }
  \label{tab:f_score}
  \scriptsize%
  \centering%
  \begin{tabu}{%
  	  rr%
  	  	*{4}{c}%
  	}
  	\toprule
  	   &  & Vortex & Ethanediol & Combustion & Isotropic   \\
  	\midrule
        \multirow{3}{*}{FPR} 
            & UP-2  & \textbf{0.122} & \textbf{0.034} & \textbf{0.476} & \textbf{0.184} \\
            & UP-5  & 0.210 & 0.044 & 0.628 & 0.239 \\
            & UP-10  & 0.375 & 0.056 & 0.759 & 0.324 \\
            & RA-UA  & 1.000 & 0.212 & 0.985 & 0.386 \\
            & RA     & 1.000 & 0.334 & 0.999 & 0.842  \\
  	\midrule
        \multirow{3}{*}{FNR} 
        & UP-2     & 0.042 & 0.036 & 0.003 & 0.003 \\
        & UP-5     & $0.8 \times 10^{-3}$ & $10^{-4}$ & $0.7 \times 10^{-4}$ & $0.96 \times 10^{-5}$ \\
        & UP-10    & \textbf{0.0} & \textbf{0.0} & $1.6 \times 10^{-7}$ & \textbf{0.0} \\
        & RA-UA  & \textbf{0.0} & \textbf{0.0} & \textbf{0.0} & \textbf{0.0} \\
        & RA     & \textbf{0.0} & \textbf{0.0} & \textbf{0.0} & \textbf{0.0}  \\
  	\bottomrule
  \end{tabu}%
\end{table}

\begin{figure}[tbph]
  \centering 
  \includegraphics[width=\columnwidth]{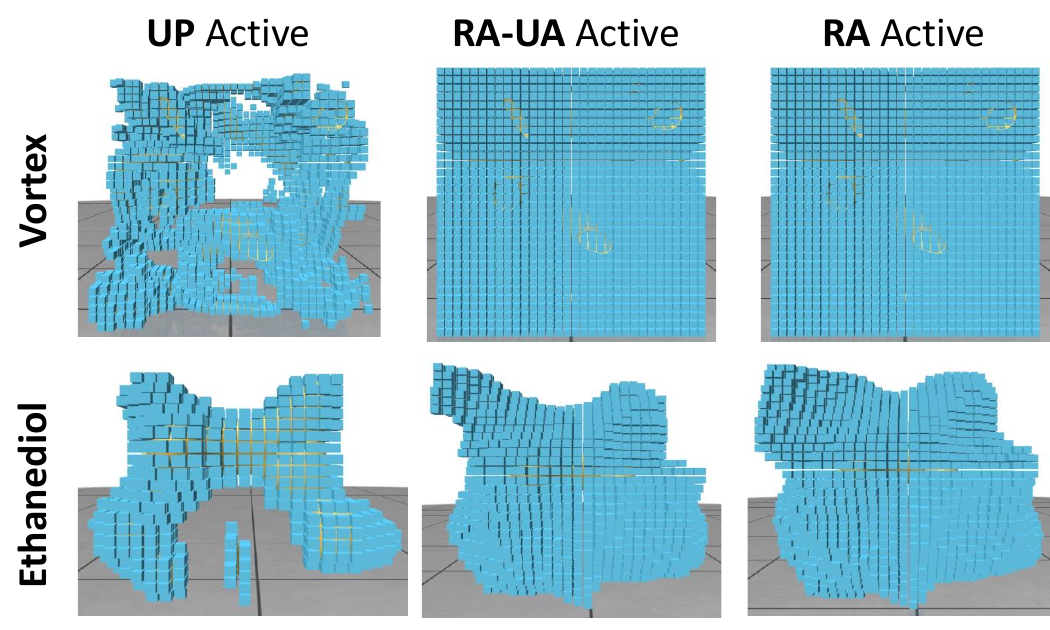}
  \caption{%
  Predicted active cells from different approaches. Our approach UP has the most accurate prediction. The other two baselines tend to overpredict the number of active cells. 
  }
  \label{fig:active}
\end{figure}

\clrb{
The false positive rate (FPR) of the uncertainty propagation method is less than the FPR of the two baselines, which explains why our approach can achieve a much shorter iso-surface extraction time. }
We also render the predicted active cells from each method in \cref{fig:active} to demonstrate the false positive rate. Ideally, active cells should only be the cells that have iso-surface crossing. Observing the predicted active cells by RA and RA-UA in the second and the third column in \cref{fig:active}, in a lot of cases, these two methods predict non-active cells as active cells. Our approach predicts a lot fewer cells as active, which are tightly distributed near the iso-surface. Therefore, the false positive rate is lower.

\clrb{
In terms of the false negative rate (FNR), because the analysis bounds are guaranteed to include the real bounds, vanilla range analysis will have zero false negative. Imperfect iso-surface extraction is possible for our uncertainty propagation approach, which is confirmed by the FNR reported in \cref{tab:f_score}. However, the FNR of the uncertainty propagation approach is very small across all datasets and controllable by the chosen value $t$.
}

\begin{figure}[tbph]
  \centering 
  \includegraphics[width=\columnwidth]{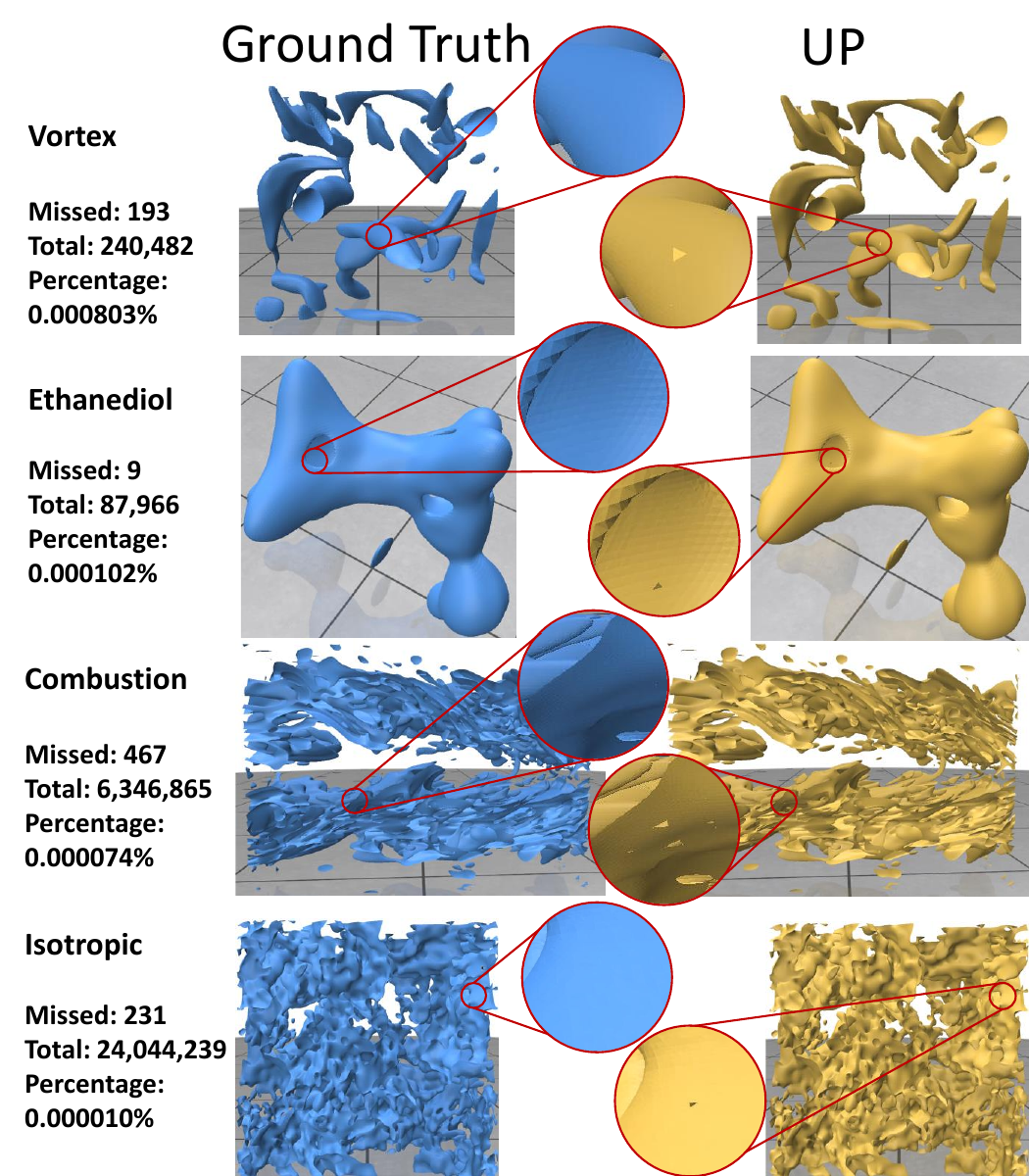}
  \caption{%
  	Comparison of the extracted iso-surfaces. The visual differences between our approach UP and the ground truth are highlighted by the circles. Our approach achieves better efficiency at the expense of only small errors.
  }
  \label{fig:qualitative}
\end{figure}

\clrb{
We also analyze the missing iso-surface components qualitatively and show the results in \cref{fig:qualitative}. The first column shows the ground truth iso-surfaces extracted by dense data reconstruction from INRs. The second column shows the iso-surfaces extracted using uncertainty propagation. We highlight the subtle missing components with red circles. 
}

\subsection{Scalability} 
\begin{figure}[tb]
  \centering 
  \includegraphics[width=\columnwidth]{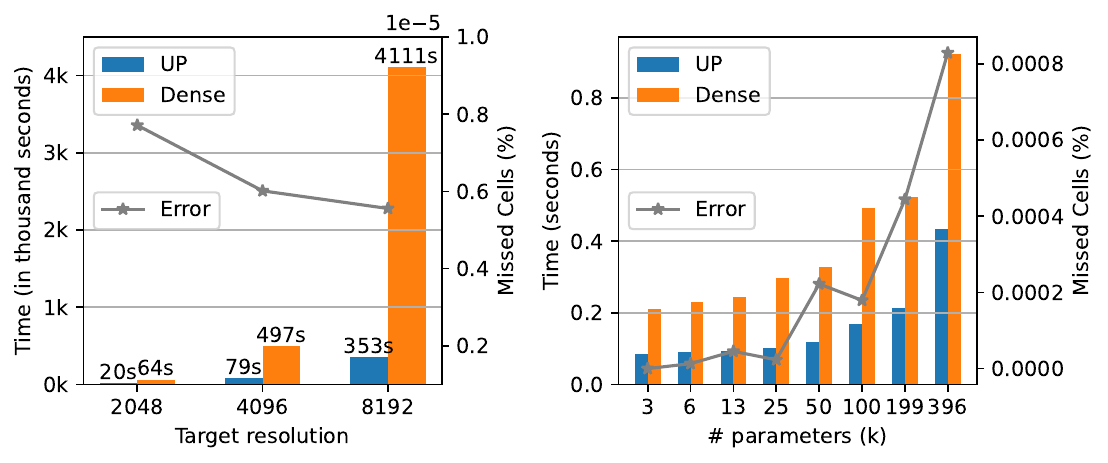}
  \caption{%
    The iso-surface extraction time and quality comparison for higher resolutions (left) and different network sizes (right). 
    The bar graphs show the extraction time comparison, while the line graphs present the missed cell percentage of our approach.  } 
  \label{fig:scal}
\end{figure}

To show the computational efficiency of our approach, we test its scalability on INRs to higher resolutions and larger networks. When increasing the reconstruction granularity of the INR, the network parameters and the training data are kept the same. The isotropic dataset is used in the resolution scalability test.
We test the iso-surface extraction time and quality in $2048^3$, $4096^3$, and $8192^3$ resolution and show the results in \cref{fig:scal}.
We observe that the time for extraction increases cubically with the increase of the resolution in the dense reconstruction case, while the extraction time for our method increases quadratically. This aligns with our expectations because we are dealing with iso-surfaces in 3D volumes. The number of reconstruction grids in the dense case increases cubically. The number of reconstructions needed in the hierarchical extraction case increases with the number of active cells which is roughly quadratic to the resolution.
The extraction error reduces slightly with the increase of resolution. This may be because distribution estimation is more accurate in cells of smaller sizes.

In terms of the network size, we tested the network with parameters ranging from 3k to 400k, which covers the sizes typically used in INR network \cite{weiss2022fast, lu2021compressive}. 
\clrb{
We tweak the network width to change the total number of parameters but keep the number of layers untouched.
}
We trained these networks on the Ethanediol dataset and tested on the target resolution of $256^3$. The extraction time and quality of our approach and dense reconstruction are shown in \cref{fig:scal}. We observe that our approach is consistently faster with decent accuracy.
However, the active cell prediction error increases with the increase in the network size. This may be due to the fact that larger networks are less likely to satisfy the CLT assumptions and the distribution estimates are less accurate. In practice, we still get good extraction results ($< 0.001\%$ error) on our largest tested network. For even larger networks, using a larger threshold $t$ could potentially solve this issue.

\subsection{Evaluation of the Estimated Distribution}

The reason for the higher efficiency of our approach lies in an accurate estimation of the INR output distribution. 
\clrb{To get the true distribution, we perform Monte Carlo sampling over an input region.
Inside a specific 3D block, we randomly sample $10^6$ positions following a uniform distribution across the block and infer the INR to get the corresponding scalar values at the sample locations.}
A histogram is built using the output values representing the ground truth INR output distribution over the specific region. 
\clrb{We compare the estimated distribution from our approach to the true distribution and two baselines. The first is RA-UA introduced in \cref{sect:RA-UA} and the second is Gaussian distribution estimate from a limited number of samples.
Because our approach is approximately $100\times$ more expensive than a regular scale value inference from the INR in the experiment, we choose $100$ samples in this baseline to match the computation cost and use these samples to fit a Gaussian distribution. We refer to this baseline as ``SAMPLE'' in the experiment results.
}
\begin{figure}[htbp]
  \centering 
  \includegraphics[width=\columnwidth]{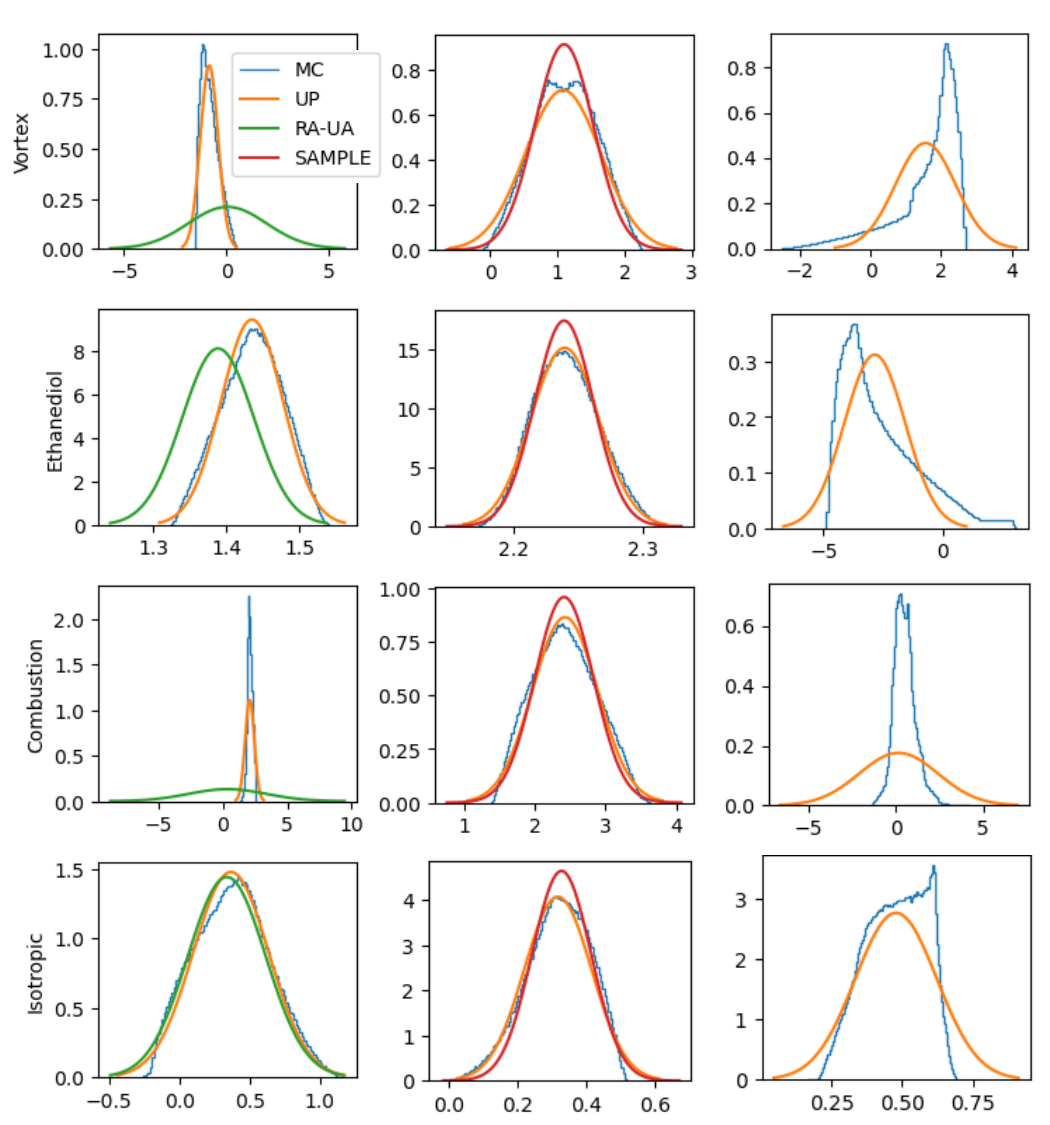}
  \caption{%
  \clrb{The comparison between the estimated output distributions versus the true data distribution obtained by Monte Carlo sampling (MC). Different rows show the results from different experimented datasets. In the first column, we compare UP with RA-UA. In the second column, we compare UP with the Gaussian estimate using 100 samples. In the third column, we select some cases where the estimation using UP is inaccurate.}
  }
  \label{fig:dist}
\end{figure}
\clrb{The estimated Gaussian distributions from different approaches and the true data distribution are plotted in \cref{fig:dist}. 
Different rows represent different datasets.}
The first column compares UP and RA-UA, although the true distribution is not necessarily Gaussian in every region, our approach reasonably approximates this distribution. Compared to the RA-UA, which predicts a biased mean and too large variance, our approximation has a much more similar distribution to the histogram obtained by Monte Carlo sampling.
\clrb{In the second column, we compare UP with the Gaussian distribution estimated using 100 samples (SAMPLE). Our approach is slightly closer to the true distribution and the Gaussian estimate using 100 samples is also reasonably accurate. This indicates that we may use the Gaussian distribution estimated from a limited number of samples to generate the soft bound and then perform hierarchical iso-surface extraction. However, more experiments are needed to decide how many samples are necessary.
In the last row, we show some cases where our estimate deviates from the ground truth. When the original data is not Gaussian distributed (rows 1 and 2), the true data distribution can exceed the predicted soft bounds, which will lead to false negatives in active cell prediction. We also identify some cases (rows 3 and 4) where the predicted variance is much larger than the true variance, which could lead to false positives in the prediction.
}

\begin{table}[bt]
  \caption{%
  \clrb{The average KL divergence of the Monte Carlo histogram from the estimated Gaussian distribution using UP, RA-UA, and limited samples. The divergence is calculated in the natural unit of information (nat).}
  }
  \label{tab:kl}
  \scriptsize%
  \centering%
  \begin{tabu}{%
  	  r%
  	  	*{4}{c}%
  	}
  	\toprule
  	$D_\text{KL}$ (nat) & Vortex & Ethanediol & Combustion & Isotropic   \\
  	\midrule
  	UP  & \textbf{0.203} & \textbf{0.048} & 0.497 & \textbf{0.126} \\
        RA-UA  & 1.563 & 2.104 & 3.173 &  1.218 \\
        SAMPLE & 0.244 & 0.044 & \textbf{0.482} & 0.129 \\
  	\bottomrule
  \end{tabu}%
\end{table}

Quantitatively, we calculate the KL divergence of the Monte Carlo histogram $P$ from the estimated Gaussian distribution $Q$. The KL divergence is calculated using the following equation:
\begin{equation}
    \infdiv{P}{Q} = \int_{-\infty}^{\infty} p(x)\log{\frac{p(x)}{q(x)}} \dif{x}.
\end{equation}
The KL divergence (the lower the better) of $P$ from $Q$ can be interpreted as the expected excess surprise from using $Q$ as a model when the real distribution is $P$.
We average the KL divergence calculated from 100 randomly sampled blocks for different datasets and show the results in \cref{tab:kl}.
\clrb{
Our method has significantly lower KL divergence compared to RA-UA and slightly lower KL divergence compared to the limited sample method, which confirms the qualitative observations.
We also find that the KL divergences of different datasets match our iso-surface extraction time and quality. 
If the KL divergence is lower for a dataset, the scalar values are more likely to be Gaussian-distributed, and our approach works better, i.e., has lower computation time and higher extraction quality.
}

\section{Discussion and Future Works}
\label{sect:discussion}
There are a few potential issues and limitations not fully discussed in the methods and the evaluation. We will discuss them and our future work in this section. 
\clrb{
Our work assumes that data values are Gaussian distribution in any arbitrary region. Although this assumption cannot be guaranteed, Gaussian-distributed data is very common in scientific simulations. If the original field follows this Gaussian assumption, its implicit neural representation should also follow the assumption as long as we have a reasonable fitting accuracy.
Unless we have other prior knowledge about the data distribution, Gaussian distribution is a natural choice that has nice properties for calculation and descent accuracy to fit the true data distribution.
From another perspective, we find that the affine form of the approximation to the INR output is also likely to be Gaussian distributed if the CLT conditions hold. Even though we cannot strictly prove the satisfaction of the CLT conditions, we find they are likely to be satisfied except for some special conditions. These three conditions are discussed in detail next. Because the observations from the two perspectives meet, we choose to model the regional INR output distribution as Gaussian in our method.
}

\textit{Large number assumption.} 
The number of random variables in PAF is equal to the number of activation functions in the last layer plus the input degree of freedom, which means the large number assumption holds in most cases. However, in the first few layers of INR, the number of random variables could be small. This can lead to inaccurate distribution estimation in the first few layers, and these errors can be carried to the output distribution.

\textit{Independence assumption.} 
We also assume each of the errors and input uncertainties are independent. However, they might not be fully independent in the application of uncertainty propagation on INRs. 
Introduced approximation errors and the input uncertainties could have complex dependencies between each other. 

\textit{Lindeberg's Condition.} The last assumption for the central limit theorem is that the random variables are similarly distributed. In the classical CLT, it is even required that each random variable is identically distributed. This requirement can be relaxed to variables with similar distributions that satisfy Lyapunov's or Lindeberg's condition. Intuitively, these conditions ensure that no random variable has a significantly large deviation from the mean. Unfortunately, it is not possible to show the satisfaction of any of these two conditions on an arbitrary INR, because no restriction is given to the weights when the network is trained. 
\clrb{However, in practice, because we minimize the mean squared error to every activation function, the variance of each uncertainty unit in the PAF is similarly small.}

\clrb{In summary, we apply the CLT and model the INR output value as Gaussian distributions to increase the estimation efficiency. In theory, these assumptions may fail in some special cases, which leads to errors in the estimation. However, because the real-world data are likely to follow Gaussian distributions in most regions, in the experiments, we find the error is small enough to provide a decent estimation so that the efficiency of the iso-surface extraction is increased.}

The major limitation of our technique is that we cannot guarantee the correct extraction result. We show the number of incorrect active cells is small. However, these errors can have unexpected consequences, for example, changing the topology of the extracted iso-surface. Our future studies aim to improve the approach to ensure correct topology or having bounded errors for the extracted iso-surfaces. However, in the current stage, we suggest using our approach for a fast preview of large INR data. The resulting iso-surfaces are more suitable for visualization than geometry or topology analysis.
Another limitation of this work is the implementation under the JAX framework. To prototype our approach and perform a fair comparison to the baseline methods, we follow the framework used by Sharp and Jacobson \cite{sharp2022spelunking}. Only simple MLP-based INR is implemented in the experiments. Although extension to other architectures should be straightforward, this implementation still limits the easy adoption of popularly used frameworks like PyTorch and TensorFlow.


\clrb{We show some results of using our approach on other types of data like SDFs and other geometric query tasks like ray casting in the appendix. Our approach works equally well on mesh extraction from SDFs as from scientific scalar fields. 
Regarding the ray casting results, our approach is more efficient than RA-Full but much less efficient than RA-Fixed. This observation is also backed by the experiments in the range analysis paper \cite{sharp2022spelunking}. Details about these experiments can be found in the appendix. To conclude, the uncertainty propagation method is superior to all range analysis variants in 3D geometric query tasks but less efficient than RA-Fixed in the 1D case due to the long overhead to track all the uncertainty units in the affine form.
}
In the future, we plan to adopt uncertainty propagation to other applications like direct volume rendering and distribution-based data summary on INRs. There is great potential to increase the efficiency of different visualization tasks.  

\section{Conclusion}

In this paper, we propose an uncertainty propagation technique through implicit neural representations using probabilistic affine forms, which allows an efficient hierarchical iso-surface extraction from the INR by estimating the output value bounds over a region. 
Compared to the baseline methods, our approach achieves a significant increase in extraction efficiency while preserving good extraction quality. The proposed technique can also be generalized to other visualization tasks on implicit neural representations.

\section*{Acknowledgments}
This work is supported in part by the National Science Foundation Division of Information and Intelligent Systems-1955764, the National Science Foundation Office of Advanced Cyberinfrastructure-2112606, U.S. Department of Energy Los Alamos National Laboratory contract 47145, and UT-Battelle LLC contract 4000159447 program manager Margaret Lentz.


{\appendices
\section{Ray Casting on SDFs}
The uncertainty propagation approach proposed in this paper can also be applied to other geometric query tasks, for example, ray casting a signed distance function (SDF). In this section, we use INR-represented SDFs of two 3D models. The fox model uses the ReLU activation function with around 7558 parameters. The bunny model uses the ELU activation function with 29441 parameters.
We follow the algorithm presented in the original range analysis paper for ray casting. The resolution for the fox model is $512 \times 512$ and the resolution for the bunny model is $400 \times 400$. 
We show this application's rendering time and quality and compare our method to the range analysis. 

The experiments performed by Sharp and Jacobson \cite{sharp2022spelunking} have already shown range analysis with reduced affine coefficients is superior to the vanilla range analysis. The RA-Fixed strategy works best, where only input uncertainty units are kept and all other coefficients are combined. Therefore, we compare our approach with the vanilla range analysis (RA-Full) and RA-Fixed and show the results in \cref{fig:raycasting}.

\begin{figure}[htbp]
  \centering 
  \includegraphics[width=\columnwidth]{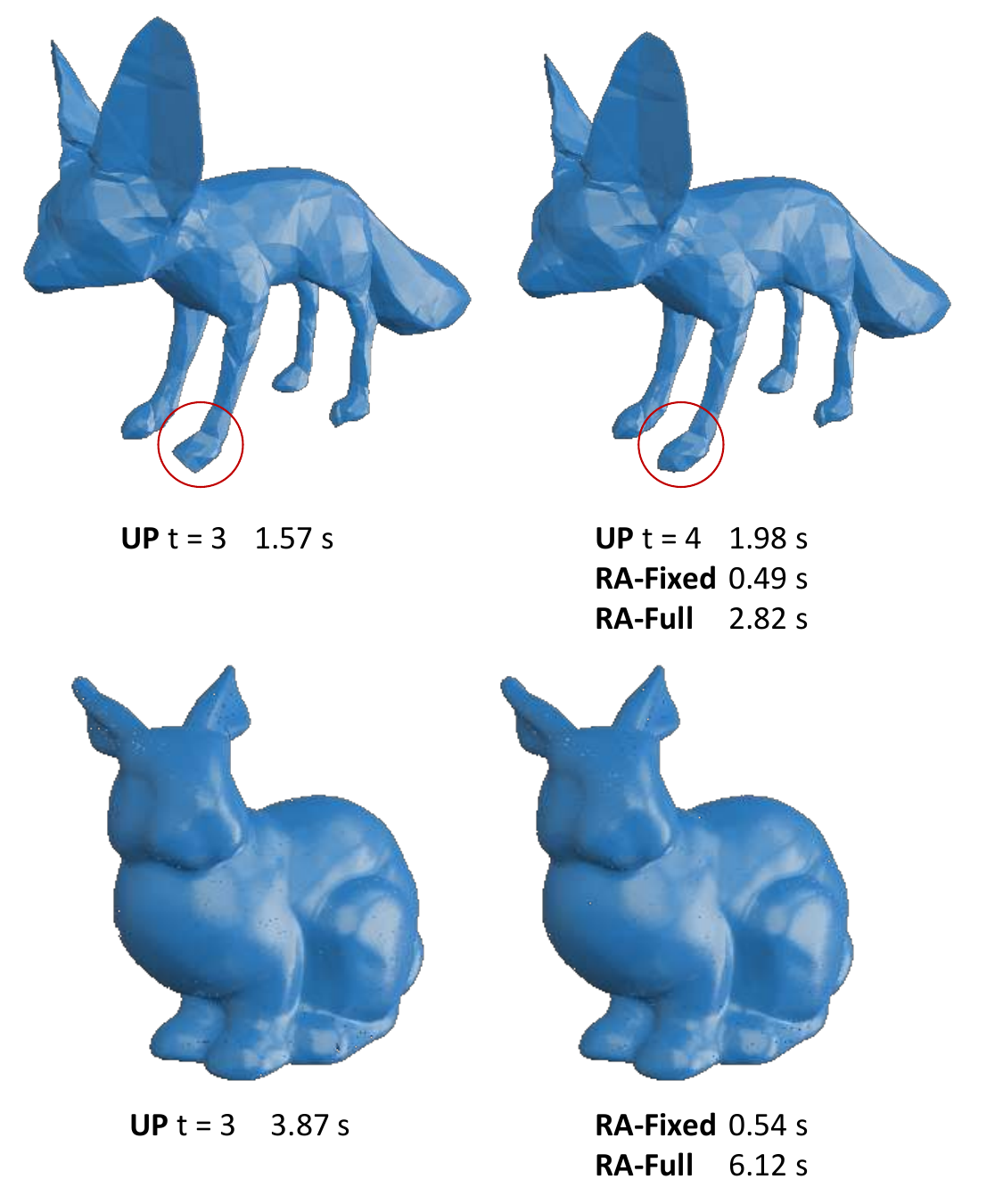}
  \caption{%
  Rendering of two models represented by signed distance functions through ray casting. The rendering time is presented under the corresponding image. Some imperfect rendering results when choosing a low threshold $t$ using our method are highlighted. 
  }
  \label{fig:raycasting}
\end{figure}

For the fox model, uncertainty propagation with the threshold $t\ge 4$, RA-Fixed, and RA-Full all produce similar results. Uncertainty propagation with $t<4$ can produce inaccurate results with missing parts in the rendering as highlighted in \cref{fig:raycasting}. Regarding the rendering time, we confirm that RA-Fixed is much more efficient than RA-Full in the ray casting task. The uncertainty propagation is more efficient than RA-Full but takes more time than RA-Fixed. The results show that the uncertainty propagation is not efficient enough for 1D geometric query tasks like ray casting, due to the overhead to track all the uncertainty units in the affine form. However, our proposed approach is more efficient in 3D geometric query tasks like iso-surface extraction.

\section{Mesh Extraction from SDFs}

In addition to the scientific scalar fields presented in the paper, we perform experiments to extract mesh from SDFs using uncertainty propagation and compare that to the range analysis. Hierarchical mesh extraction from SDFs is essentially the same as our hierarchical iso-surface extraction using iso-value 0. The computation time and quality (measured in false negative rate and false positive rate) are presented in \cref{tab:mesh_extraction}.

\begin{table}[htbp]
  \caption{%
    We present the mesh extraction time and quality using uncertainty propagation (UP) and range analysis (RA) for two datasets. ``Dense Time'' denotes the baseline extraction time if we reconstruct the data in a dense grid.
    }
  \label{tab:mesh_extraction}
  \scriptsize%
  \centering%
  \begin{tabu}{%
  	  r%
  	  	*{5}{c}%
  	}
  	\toprule
  	Data & Dense Time & Method & Time & FNR & FPR    \\
  	\midrule
        \multirow{2}{*}{Fox} & \multirow{2}{*}{0.253s} & UP    & 0.088s & 0.0 & 0.035  \\
                             & & RA     & 0.112s & 0.0 & 0.072   \\
  	\midrule
        \multirow{2}{*}{Bunny} & \multirow{2}{*}{0.333s} & UP     & 0.125s & 0.0 & 0.063 \\
                          &  & RA     &  0.187s & 0.0 & 0.121   \\
  	\bottomrule
  \end{tabu}%
\end{table}

Uncertainty propagation takes less computation time than range analysis in mesh extraction and both of them are more efficient than dense reconstruction. When experimenting on scientific scalar fields, range analysis is less efficient than dense reconstruction in most cases. Since the INR structure and the algorithm are the same for SDFs and scalar fields, the only difference is the datasets themselves. The complex and high-frequency features in the scientific scalar fields may lead to complex INR models and conservative bounds from range analysis, which motivates the use of uncertainty propagation. 

Zero false negative rate in this experiment indicates that uncertainty propagation does not miss any components of the surface under the threshold of choice $t=5$. A lower false positive rate of uncertainty propagation compared to range analysis shows that the lower computation time is due to more accurate bound estimation.

\section{Analysis of the INR Quality}

In the paper, we present the INR size compared to the original uncompressed data size and the PSNR of the reconstructed volume. In this section of the appendix, we provide more information on the dataset and the qualitative analysis of the INR representation. Although the INR quality is not the focus, it serves as an important background that motivates this paper. 

\begin{figure}[htbp]
  \centering 
  \includegraphics[width=\columnwidth]{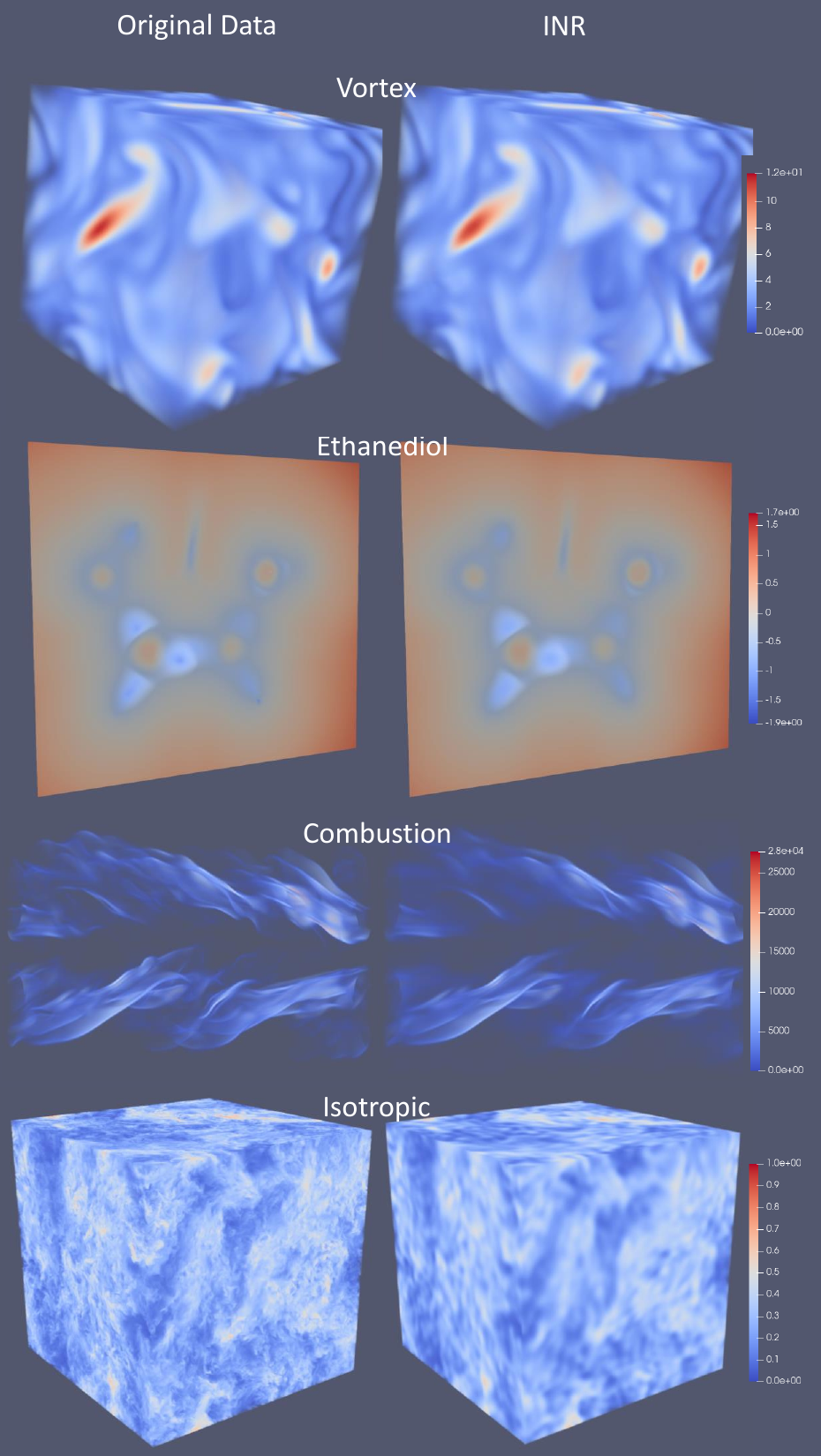}
  \caption{%
  The volume rendering images for four tested datasets. The renderings of the original data are shown in the left column, while the renderings of the INR reconstructions are shown in the right.
  For the ethanediol dataset, we show a slice of the volume along with the volume rendering depicting the Ethandiol structure.
  }
  \label{fig:volume}
\end{figure}

Comparing the INR reconstruction results with the original data, we find the reconstruction quality is very high for the vortex and the ethanediol dataset, which is also confirmed by the high PSNR values in the paper. Regarding the combustion and the isotropic dataset, the INR reconstruction blurs some regions of the original data due to a limited number of parameters in the network. Overall, we can see scientific datasets often contain more complex features compared to signed distance functions for 3D models. This may explain why the range analysis approach performs poorly on INRs trained on scientific datasets. The range analysis outperforms the dense reconstruction only on the ethanediol dataset, which is the only relatively simple dataset. It is worth noting the poor performance of range analysis is also related to the activation function choice, which we will show next.

\section{Solution to Activation Function Approximations}

In section IV-B of the paper, we explain that the probabilistic affine forms can be propagated through the nonlinear activation functions by approximating them with linear functions. The linear approximation can be solved analytically using the least square method according to eq. (10). We show the solution to three commonly used activation functions, Sine, ReLU, and ELU function in \cref{tab:least_square}.

\begin{table*}[htbp]
  \caption{%
  The table shows the linear approximation coefficients to different activation functions obtained by solving eq. (10) using the least square method. $\gamma^2$ is the mean squared error of this approximation. Here we use $\mu$ and $\sigma$ to denote the mean and variance of the estimated Gaussian PDF for the input probabilistic affine form. 
  The $\erf$ in the table denotes the Gauss error function.
  }
  \label{tab:least_square}
  \scriptsize%
  \centering%
  \begin{tabu}{%
  	  	*{4}{c}%
  	}
  	\toprule
  Activation & $\alpha$ & $\beta$ & $\gamma^2$  \\
  	\midrule
  ReLU & 
  \makecell{
    $a = \erf{(\frac{-\mu}{\sqrt{2} \sigma})}$
    \\ 
  $\alpha = \frac{1}{2}(1- a)$
  }
  &\makecell{
  $\beta = \frac{\sigma}{\sqrt{2\pi}}e^{\frac{-\mu^2}{2\sigma^2}}$}
  &
  \makecell{
  $\gamma^2 = \frac{1}{4}(1-a^2) (\mu^2+\sigma^2) + a \mu \beta - \beta ^ 2$}
  \\
  \midrule
  ELU & 
  \makecell{
    $a = \erf{(\frac{-\mu}{\sqrt{2} \sigma})}$
    \\
    $b = [1+\erf{(\frac{-\mu-\sigma^2}{\sqrt{2} \sigma})}] e ^ {\mu + \frac{\sigma^2}{2}}$
    \\
    $\alpha = \frac{1}{2}(1-a+b)$
  }
  &\makecell{
    $c = \frac{1}{\sqrt{2\pi}}e^{\frac{-\mu^2}{2\sigma^2}}$
    \\
    $\beta = \frac{1}{2} b (1-\mu)-\frac{1}{2}(1+a) + c \sigma$
  }
  &
  \makecell{
  $
  \gamma^2 = \frac{1}{4}(1-a^2) (\mu^2+\sigma^2) + (\frac{1}{2} ab - c^2 -\frac{1}{4}b^2-\frac{1}{2}b) \sigma^2 + ac\mu \sigma + 
  $
  \\
  $
  \;\; (a-b+1)c\sigma+\frac{1}{2}(ab-b-a^2+1)\mu + \frac{1}{4}[1+\erf{(\frac{-\mu-2\sigma^2}{\sqrt{2} \sigma})}] e ^ {2\mu + 2\sigma^2} + 
  $
  \\
  $
  \;\; \frac{1}{4} - \frac{1}{4}(a-b)^2 - \frac{1}{2}b
  $
  }
  \\
  \midrule
  Sine & 
  \makecell{
    $a = e^{-\frac{1}{2}\sigma^2}$
    \\
    $\alpha = a\cos{\mu} $
  }
  &\makecell{
    $\beta = a(\sin{\mu} - \mu \cos{\mu})$
  }
  &
  \makecell{
  $
  \gamma^2 = -a^2(\sigma ^ 2\cos^2{\mu} +\sin^2{\mu})+\frac{1}{2}[1-a^4\cos{(2\mu)}]
  $
  }
  \\
  	\bottomrule
  \end{tabu}%
\end{table*}

\section{Evaluation with Other Activation Functions}

In the evaluation sections of the paper, we only present the INR models with sine activation functions because of the representation quality. Here, we include more experiment results using other activation functions on the vortex and ethanediol dataset with the same number of parameters in the network. Similarly, we apply uncertainty propagation along with range analysis and report the computation time, the false negative rate, and the false positive rate. We also report the fitting accuracy of the INR model in PSNR.

\begin{table}[htbp]
  \caption{%
    This table presents the iso-surface extraction time and quality from INRs with different activation functions on the vortex and the ethanediol datasets. The quality is measured in false negative rate (FNR) and false positive rate (FPR). We compare uncertainty propagation (UP) and range analysis (RA) in the table. The fitting quality (PSNR) with different activation functions is also shown in the table.
    }
  \label{tab:activation_functions}
  \scriptsize%
  \centering%
  \begin{tabu}{%
  	  r%
  	  	*{6}{c}%
  	}
  	\toprule
  	Data & Activation &PSNR & Method & Time & FNR & FPR    \\
  	\midrule
        \multirow{7}{*}{Vortex} 
                             & \multirow{2}{*}{Sine} & \multirow{2}{*}{38.23} 
                             & UP    & 0.176s & 0.0008 & 0.210  \\
                             & & & RA    & 0.386s & 0.0 & 1.0  \\
                             \cmidrule{2-7}
                             & \multirow{2}{*}{ReLU} & \multirow{2}{*}{31.31} 
                             & UP    & 0.154s & 0.001 & 0.198  \\
                             & & & RA & 0.299s & 0.0 & 0.706  \\
                             \cmidrule{2-7}
                             & \multirow{2}{*}{ELU} & \multirow{2}{*}{30.54} 
                             & UP    & 0.154s & 0.0003 & 0.172  \\
                             & & & RA    & 0.250s & 0.0 & 0.482  \\
  	\midrule
        \multirow{7}{*}{Ethanediol} & \multirow{2}{*}{Sine} & \multirow{2}{*}{54.87} 
                                & UP    & 0.099s & 0.0001 & 0.044  \\
                             & & & RA    & 0.235s & 0.0 & 0.334  \\
                             \cmidrule{2-7}
                             & \multirow{2}{*}{ReLU} & \multirow{2}{*}{51.28} 
                             & UP & 0.091s & 0.012 & 0.039  \\
                             & & & RA    & 0.120s & 0.0 & 0.078  \\
                             \cmidrule{2-7}
                             & \multirow{2}{*}{ELU} & \multirow{2}{*}{49.45} 
                             & UP    & 0.096s & 0.00007 & 0.045  \\
                             & & & RA    & 0.121s & 0.0 & 0.087  \\
  	\bottomrule
  \end{tabu}%
\end{table}

Range analysis is more efficient on ReLU and ELU functions compared to the Sine function. This is possibly due to that the approximation method used by range analysis has a high error on periodic functions and thus has conservative bounds.
Uncertainty propagation is more efficient than range analysis under all activation functions as shown in \cref{tab:activation_functions} and the benefit is most significant for sine functions.  
Because the network size and the extraction resolution is kept the same, the dense reconstruction time for all settings is around 0.25 seconds. 
The computation time for the ethanediol dataset is much less compared to the vortex dataset using RA and UP because the ethanediol dataset has fewer high-frequency features compared to the vortex dataset and the total area of the tested iso-surface is smaller according to \cref{fig:volume}.
Based on these observations and the experiments of mesh extraction from SDFs, we can conclude that the poor performance of range analysis in our experiments is due to both the data complexity and the choice of activation functions. Uncertainty propagation is always more efficient with very small errors in the iso-surface extraction task.
}

\bibliographystyle{IEEEtran}
\bibliography{main.bib}

\vspace{11pt}

\begin{IEEEbiography}[{\includegraphics[width=1in,height=1.25in,clip,keepaspectratio]{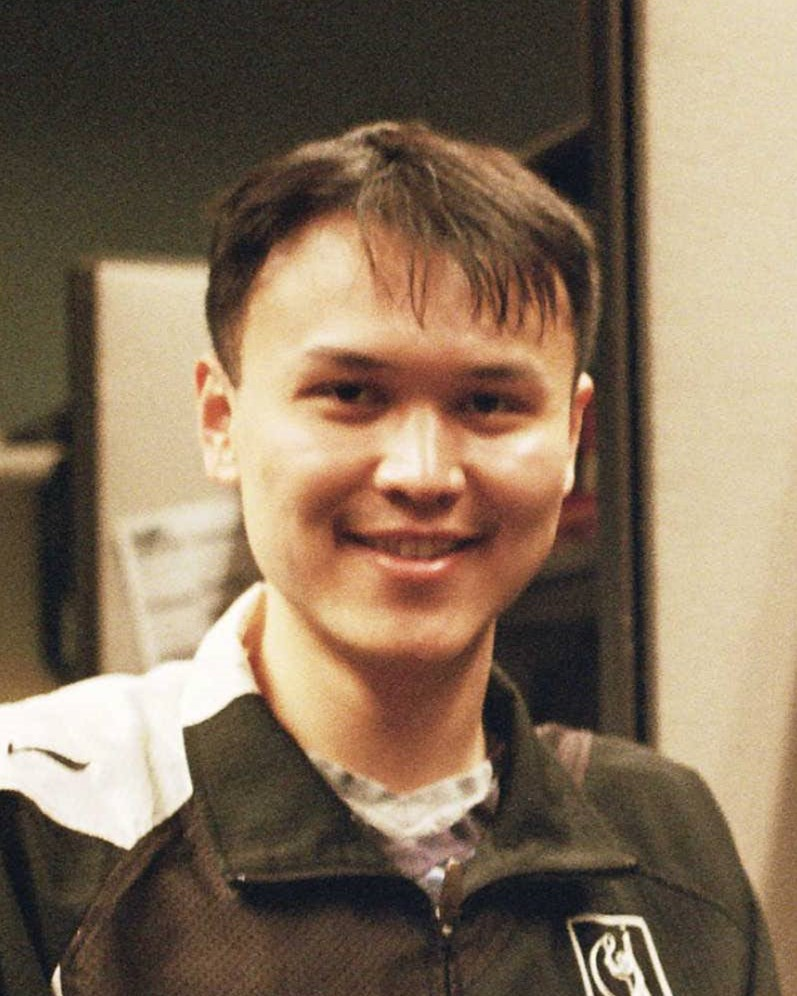}}]{Haoyu Li}
is a Ph.D. student in the Department of Computer Science and Engineering at the Ohio State University. He received his B.S. degree in Psychology from Beijing Normal University in 2017. His research interests are mainly in visualization for particle simulations and machine learning for scientific visualization.
\end{IEEEbiography}

\vspace{11pt}
\begin{IEEEbiography}[{\includegraphics[width=1in,height=1.25in,clip,keepaspectratio]{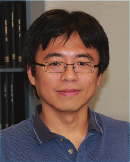}}]{Han-Wei Shen}
is a full professor at the Ohio State University. He received his B.S. degree from the Department of Computer Science and Information Engineering at National Taiwan University in 1988, his M.S. degree in computer science from the State University of New York at Stony Brook in 1992, and his Ph.D. degree in computer science from the University of Utah in 1998. From 1996 to 1999, he was a research scientist at NASA Ames Research Center in Mountain View California. His primary research interests are scientific visualization and computer graphics. He is a winner of the National Science Foundation CAREER award and the U.S. Department of Energy Early Career Principal Investigator Award. He also won the Outstanding Teaching award twice in the Department of Computer Science and Engineering at the Ohio State University.
\end{IEEEbiography}

\clearpage

\vfill

\end{document}